%% file: XPU.tex
\documentclass[sigplan, 10pt]{acmart}
\makeatletter
\def\@ACM@checkaffil{% Only warnings
    \if@ACM@instpresent\else
    \ClassWarningNoLine{\@classname}{No institution present for an affiliation}%
    \fi
    \if@ACM@citypresent\else
    \ClassWarningNoLine{\@classname}{No city present for an affiliation}%
    \fi
    \if@ACM@countrypresent\else
        \ClassWarningNoLine{\@classname}{No country present for an affiliation}%
    \fi
}
\makeatother

\usepackage{booktabs}
\usepackage{bbding}

\usepackage{tikz}
\usepackage{amsmath}

\usepackage{filecontents}
\usepackage[normalem]{ulem}
\usepackage{graphicx} % Required for inserting images
\usepackage{pifont}
\usepackage{balance}
\usepackage{multirow}
\usepackage{enumitem}
\usepackage{lipsum}
\usepackage{xspace}
\usepackage[]{hyperref}
\usepackage{amsmath}
\usepackage{algorithm}
\usepackage{algpseudocode}

\usepackage{multirow}
\usepackage{listings}
\usepackage{xcolor}
\usepackage{makecell}
\usepackage{tikz}

\definecolor{deepblue}{RGB}{0, 0, 139}
\newcommand{\red}{\color[HTML]{FE0000}}%
\newcommand{\green}{\color[HTML]{008000}}%

\usepackage{titlesec}
\titlespacing*{\section}{0pt}{*0.9}{*0.9}
\titlespacing*{\subsection}{0pt}{*0.9}{*0.9}
\titlespacing*{\subsubsection}{0pt}{*0.9}{*0.9}

\definecolor{deepYellow}{HTML}{FF8C00}  % (Dark Orange)
\definecolor{deepGreen}{HTML}{228B22}   % (Forest Green)

\newcommand{\FDH}[1]{\textcolor{black}{#1}}
\newcommand{\CL}[1]{\textcolor{black}{#1}}
\newcommand{\FEH}[1]{\textcolor{black}{#1}}
\newcommand{\FEHSOSP}[1]{\textcolor{black}{#1}}
\newcommand{\CLSOSP}[1]{\textcolor{black}{#1}}

\newcommand{\myparagraph}[1]{\noindent\textbf{\emph{#1}}}
\newcommand{\sys}{HeteroInfer\xspace} %need to include \usepackage{xspace}
\newcommand{\htl}{Hetero-layer\xspace}
\newcommand{\htt}{Hetero-tensor\xspace}

\newenvironment{myitemize}%
  {\begin{list}{\labelitemi}{\itemsep1pt \topsep2pt \parsep0.00in
  \partopsep=0pt \leftmargin1em}}%
  {\end{list}}

\begin{document}
%-------------------------------------------------------------------------------

%don't want date printed
\date{}

\title{Characterizing Mobile SoC for Accelerating Heterogeneous LLM Inference}

\author{Le Chen$^{1\dagger}$, Dahu Feng$^{1\S}$, Erhu Feng\textsuperscript{\Envelope}$^{\dagger}$, Yingrui Wang$^{\ddag}$, Rong Zhao$^{\S}$,
\\ Yubin Xia$^{\dagger}$, Pinjie Xu$^{2\ddag}$, Haibo Chen$^{\dagger}$}
% {cen-le@sjtu.edu.cn, fengdh21@mails.tsinghua.edu.cn, fengerhu1@sjtu.edu.cn} \\
% {wangyingrui@sensetime.com, r_zhao@tsinghua.edu.cn, xiayubin@sjtu.edu.cn} \\
% {haibochen@sjtu.edu.cn, xupinjie321@outlook.com} \\
\affiliation{
{\normalsize \it
{$^\dagger$Institute of Parallel and Distributed Systems, Shanghai Jiao Tong University}} \\
{\normalsize \it {$^\S$Tsinghua University}} 
{\normalsize \it {$^\ddag$SenseTime Research}} \\
}

\renewcommand{\authors}{Le Chen, Dahu Feng, Erhu Feng, Yingrui Wang, Rong Zhao, Yubin Xia, Pinjie Xu, Haibo Chen}
\renewcommand{\shortauthors}{Le Chen, Dahu Feng, Erhu Feng, Yingrui Wang, Rong Zhao, Yubin Xia, Pinjie Xu, Haibo Chen}

\acmYear{2025}\copyrightyear{2025}
\setcopyright{acmlicensed}
\acmConference[SOSP '25]{ACM SIGOPS 31st Symposium on Operating Systems Principles}{October 13--16, 2025}{Seoul, Republic of Korea}
\acmBooktitle{ACM SIGOPS 31st Symposium on Operating Systems Principles (SOSP '25), October 13--16, 2025, Seoul, Republic of Korea}
\acmDOI{10.1145/3731569.3764808}
\acmISBN{979-8-4007-1870-0/25/10}




%-------------------------------------------------------------------------------

\pagestyle{empty}

\settopmatter{printfolios=false,printacmref=true}

\input{./abs}

\maketitle

\footnotetext[1]{The two authors contributed equally to this work and should be considered co-first authors.}
\footnotetext[2]{Pinjie Xu is Project Leader.}

\input{./intro}
\input{./back}
\input{./charac}
\input{./design}
% \input{./impl}
\input{./eval}

\input{./discuss}
\input{./concl}

\begin{acks}
We sincerely thank our shepherd and the anonymous reviewers of SOSP 2025,
whose reviews, feedbacks, and suggestions have significantly strengthened our work.
We also thank SenseTime for supporting this work.
This work is supported in part by
% National Key Research \& Development Program of China (No. xxxx),
STI 2030—Major Projects 2021ZD0200300,
China National Natural Science Foundation (No. 623B2074, 62088102, 62472279),
Erhu Feng is the corresponding author.
\end{acks}

%-------------------------------------------------------------------------------
\bibliographystyle{plain}
\bibliography{references}

%%%%%%%%%%%%%%%%%%%%%%%%%%%%%%%%%%%%%%%%%%%%%%%%%%%%%%%%%%%%%%%%%%%%%%%%%%%%%%%%
\end{document}

%% file: abs.tex
\begin{abstract}

With the rapid advancement of artificial intelligence technologies such as ChatGPT, AI agents, and video generation,
contemporary mobile systems have begun integrating these AI capabilities on local devices to enhance privacy and reduce response latency. 
To meet the computational demands of AI tasks, current mobile SoCs are equipped with diverse AI accelerators, including GPUs and Neural Processing Units (NPUs). 
% However, a comprehensive characterization of these heterogeneous processors has not yet been achieved. 
% Existing designs typically leverage only a single AI accelerator for LLM inference, 
% resulting in suboptimal utilization of computational resources and memory bandwidth.
However, there has not been a comprehensive characterization of these heterogeneous processors, 
and existing designs typically only leverage a single AI accelerator for LLM inference,
% However, existing inference engines on mobile devices
%  typically utilize a single type of AI accelerator, 
leading to suboptimal use of computational resources and memory bandwidth. 
% In this paper, we present \sys, the fastest LLM inference engine for heterogeneous processors on real-world mobile devices. 

\FEHSOSP{In this paper, we first summarize key performance characteristics of heterogeneous processors, SoC memory bandwidth, etc.
Drawing on these observations, we propose different heterogeneous parallel mechanisms to fully exploit both GPU and NPU computational power and memory bandwidth.
We further design a fast synchronization mechanism between heterogeneous processors that leverages the unified memory architecture.
By employing these techniques, we present \sys, the fastest LLM inference engine in mobile devices which supports GPU-NPU heterogeneous execution. 
Evaluation shows that \sys delivers a 1.34$\times$ to 6.02$\times$ end-to-end speedup over state-of-the-art GPU-only and NPU-only LLM engines,
while maintaining negligible interference with other applications.}

\end{abstract}

%% file: intro.tex
\section{INTRODUCTION}

\FDH{Driven by the rapid evolution in large language models (LLMs)}, 
technologies such as ChatGPT~\cite{chatgpt-3.5, mcintosh2023google, Claude, glm2024chatglm}, AI agents~\cite{hong2024cogagent,you2025ferret,wang2024mobile2,wang2024mobile}, 
and video generation~\cite{yang2024cogvideox,Qwen2VL,hong2022cogvideo,zhang2024internlm} have gained widespread adoption. 
Concurrently, as users increasingly prioritize the privacy of their personal data, 
there is a growing trend towards executing model inference on local devices like smartphones. 
To enable the efficient calculation of large language models on these mobile platforms, 
contemporary mobile System-on-Chip (SoC) manufacturers have integrated various AI accelerators, including GPUs and neural processing units (NPUs). 
These accelerators~\cite{V10, xue2023vnpu, kwon2021heterogeneous, SCAR, jouppi2017datacenter, gptpu, xue2024hardware, 10.1145/3524453} enhance capabilities for vector and matrix computations, aligning with the computational demands of AI applications. 
For example, Qualcomm's SoCs incorporate Adreno GPUs~\cite{adreno-gpu} and Hexagon NPUs~\cite{hexagon-npu} to address the computing needs of edge AI applications. 
% Furthermore, by integrating different computational units within a single SoC, these processors can utilize a unified physical memory, 
% thus obviating the explicit data copying.
% between the discrete devices and  
% requirement of data transferring between the discrete GPU and CPU in desktops and servers.

Prior studies~\cite{jiang2020BytePS, hsu2023SHMT, hsu2024shmt} have proposed inference engines to leverage the computation capabilities of heterogeneous processors.
However, these solutions are predominantly tailored for cloud infrastructures and discrete accelerators,
rendering them incompatible with mobile platforms.
Although some researches~\cite{mnn,mlc-llm,xu2024fastondevicellminference} have explored mobile inference engines, 
they still fall short in fully exploiting the \FEHSOSP{computational power} of mobile heterogeneous SoCs.
The limitation stems from three main factors:

\underline{\emph{First}}, the pronounced performance disparity between mobile NPUs and GPUs.
For instance, on the Snapdragon 8 Gen 3 platform~\cite{Snapdragon8gen3}, the GPU achieves approximately 1 TFLOPS in practice (with a theoretical peak of 2.8 TFLOPS),
while the NPU delivers up to 10 TFLOPS (in actual) performance.
Thus, \CLSOSP{enforcing parallel execution of the GPU and NPU solely based on their raw computational power may not guarantee an improvement in end-to-end performance.}
\underline{\emph{Second}}, synchronization overhead between heterogeneous processors is substantial.
On mobile platforms, GPU-NPU synchronization during LLM inference can take around 400 microseconds,
which is comparable to or even exceeds the execution time of individual GPU or NPU kernels.
\CLSOSP{\underline{\emph{Third}}, the decoding phase is inherently memory-intensive.
% As it generates one token at a time, the processors must repeatedly access the full set of weight parameters while performing only a small amount of computation per step.
% This imbalance renders memory bandwidth the primary performance bottleneck.
Merely offloading computation to GPU and NPU offers limited benefits and may even degrade performance due to additional synchronization overhead.}
Consequently, designing a LLM inference engine that can efficiently coordinate all heterogeneous processors 
in real-world mobile devices remains a substantial and unresolved challenge.

After an in-depth analysis of the heterogeneous processors within mobile SoCs, 
we observe new opportunities to enhance LLM inference by
leveraging distinctive performance characteristics for heterogeneous processors.
% we have identified \CL{several distinctive characteristics} based on their hardware architecture. 
\begin{myitemize}
\item \textbf{\emph{Tensor-sensitive NPU performance.}} 
\FEHSOSP{Although NPUs can achieve superior performance under ideal conditions, 
their efficiency is highly dependent on tensor characteristics such as order, size, and shape. 
\CLSOSP{If the characteristics of tensors involved in computation} are not well-aligned with the NPU's hardware architecture, 
both utilization and performance may degrade significantly.}

\item \textbf{\emph{Unified memory architecture across processors.}}
\CLSOSP{Unlike discrete accelerators, mobile CPUs, GPUs, and NPUs adopt a unified memory architecture (UMA) within the system memory,
featuring a shared address space that facilitates efficient inter-processor communication.}

% \item \textbf{Static NPU graph v.s. dynamic GPU execution.}
% \FEHSOSP{Existing mobile NPUs are limited to executing static computation graphs, 
% which poses challenges for handling dynamic inputs in LLMs. 
% However, mobile GPUs are more adept at accommodating dynamic workloads, 
% as each GPU kernel can manage varying tensor shapes.}

\item \textbf{\emph{Memory bandwidth improved with multiple processors.}}
A single processing unit is insufficient to fully saturate the SoC's memory bandwidth. 
For example, GPU alone can utilize only $40\sim45$ GB/s of memory bandwidth in memory-intensive workloads. 
In contrast, employing two processing units concurrently can achieve a memory bandwidth of about 60 GB/s (theoretical memory bandwidth in SoC is 68 GB/s).
\end{myitemize}

% These distinctive performance traits unlock new opportunities for LLM inference on heterogeneous processors:
% \FEHSOSP{\emph{Leveraging GPUs to improve the lower bound of NPU's fluctuant performance and enhance computational flexibility.}}

% These distinctive performance traits unlock new opportunities for enhancing GPU and NPU parallelism:
% \emph{By strategically leveraging the GPU, it can compensate for the NPU performance limitation in specific scenarios.}
% we can effectively mitigate NPU performance limitations in specific scenarios, 
% paving the way for more efficient and robust processing architectures

% \FEH{These unique performance traits present new opportunities for optimizing GPU and NPU parallelism,
% that is \emph{utilizing the GPU to compensate for the NPU performance limitation under several specific situations.}}
% which have become significant bottlenecks in mobile-side LLM inference.}

In this paper, we introduce \sys, the fastest mobile inference engine, designed to efficiently leverage all heterogeneous processing units in mobile SoCs.
The CPU is employed as a control plane for synchronization and GPU kernel scheduling, as it is ill-suited for computing tasks due to the low energy efficiency.
% Recognizing the diverse workloads on mobile platforms, 
% our approach employs the CPU as a control plane for synchronization and GPU kernel scheduling. 
The NPU serves as the primary computing unit, handling the majority of computing tasks, 
while the GPU acts as a secondary computing unit to enhance the lower bound of NPU performance.
To efficiently leverage these heterogeneous computing resources, 
\sys takes comprehensive account of the performance characteristics of the GPU and NPU, 
such as stage performance, order-sensitive performance and shape-sensitive performance.

To enable both layer-level and tensor-level GPU-NPU parallelism on real-world mobile devices, \sys further introduces three techniques.
First, \sys applies different tensor partitioning strategies during both the prefill and decoding phases to facilitate tensor-level heterogeneous execution. 
Second, \sys employs a fast synchronization mechanism based on predictable kernel waiting times to achieve microsecond-level synchronization. 
Third, \sys incorporates a tensor partitioning solver that generates optimal partitioning solutions \CLSOSP{with the help of a hardware profiler}. 
\CLSOSP{We built an industrial-grade LLM inference engine on the Snapdragon 8 Gen 3 SoC, 
one of the most advanced mobile platforms that integrates Arm CPU, GPU and NPU.
To leverage both the GPU and NPU, we developed optimized GPU kernels using OpenCL 
and integrated the NPU operators provided by Qualcomm's QNN~\cite{QNN} into our framework.}
% The GPU was efficiently utilized through optimized OpenCL operator implementations.}
% To incorporate the NPU, we integrated NPU operators provided by Qualcomm's QNN~\cite{QNN} into our framework. 
% Given that current mobile NPUs support only static computation graphs, our NPU integration is limited to the operator level. 
\CLSOSP{We avoid using the activation quantization and sparsity techniques, as these techniques may decrease the accuracy of the model inference.} 

\FEHSOSP{\sys is the first LLM engine to surpass 1000 tokens/sec in prefill phase and 50 tokens/sec in decoding phase, utilizing high-precision computation on mobile devices for billion-parameter scale LLMs.}
\FEHSOSP{In end-to-end evaluations, \sys achieves a speedup ranging from 1.34$\times$ to 6.02$\times$ compared to SOTA GPU-only and NPU-only frameworks across various ML workloads. 
During the prefill phase, \sys accelerates the speed up to 3.69$\times$ over PI-2~\cite{mlc-llm} (NPU) and 8.68$\times$ compared to MNN~\cite{mnn} (GPU). 
When the sequence length does not align with the NPU graph shape, 
\sys delivers up to a 2.12$\times$ improvement over traditional padding-based methods. 
In the decoding phase, \sys consistently outperforms existing frameworks, 
achieving speedups between 1.50$\times$ and 2.53$\times$. 
To evaluate performance interference, we run \sys concurrently with GPU-intensive workloads such as gaming. 
\sys effectively minimizes interference between LLM and gaming tasks, 
maintains stable frame rates (without FPS drop),
and incurs only a 2.2\% slowdown in the prefill phase and 17.7\% during decoding.
}
% In the prefill phase, just using the layer-level heterogeneous execution can improve the speed up to 7.27$\times$ over MLC~\cite{mlc-llm} and 3.18$\times$ over MNN~\cite{mnn}.
% Furthermore, tensor-level heterogeneous execution delivers nearly a 40\% performance improvement compared to layer-level execution.
% When the sequence length is misaligned with the shape of NPU graph, tensor-level approach achieves up to 2.12$\times$ improvement than padding approach.
% In the decoding phase, \sys still outperforms existing frameworks by 1.50$\times$ to 2.53$\times$.}
% When running concurrently with GPU-intensive workloads, \sys minimizes the interference between LLM inference and the rendering tasks (without FPS drop),
% with only a 7.26\% slowdown for LLM tasks.

%% file: back.tex
\section{BACKGROUND \& RELATED WORK}
\label{s:back}

% colors = ['#a5c2cd', '#dbc0af', '#7AA37A']
\definecolor{lightblue}{HTML}{A5C2CD}
\definecolor{lightred}{HTML}{DBC0AF}
\definecolor{lightgreen}{HTML}{7AA37A}

\newcommand{\NPU}{\textcolor{lightblue}{\tikz[baseline=0.4ex]\fill[lightblue] rectangle (2ex,2ex);}}
\newcommand{\CPU}{\textcolor{lightred}{\tikz[baseline=0.4ex]\fill[lightred] rectangle (2ex,2ex);}}
\newcommand{\GPU}{\textcolor{lightgreen}{\tikz[baseline=0.4ex]\fill[lightgreen] rectangle (2ex,2ex);}}

\begin{table*}[ht]
    \setlength{\abovecaptionskip}{5pt}
    \caption{We summarize the functionalities and limitations of current mobile-side inference engines as follows: 
	\textbf{CPU, GPU, NPU} indicate support for various backends, accommodating both integer and floating operations. 
	% \textbf{Sparsity} \FDH{refers to sparsity reliance on quantized activations or weights.} %refers to the reliance on sparsity in activations or weights. 
	% \textbf{Sparse activation} \FDH{refers to sparsity reliance on quantized activations.} %refers to the reliance on sparsity in activations or weights. 
	\textbf{Accuracy} indicates whether the model's accuracy is consistent with the original model.}
    \centering\resizebox{\textwidth}{!}{
    \label{tab:comparison}
    \begin{tabular}{c|c|c|c|c|c|c|c|c|c|c|c}
        \toprule
        \hline
        \multirow{2}{*}{\textbf{Framework}} & \multicolumn{2}{c|}{\textbf{CPU} \CPU} & \multicolumn{2}{c|}{\textbf{GPU} \GPU} & \multicolumn{2}{c|}{\textbf{NPU} \NPU} & \multirow{2}{*}{\textbf{NPU GEMM Type}} & \multirow{2}{*}{\textbf{Prefill}} & \multirow{2}{*}{\textbf{Decoding}} & \multirow{2}{*}{\textbf{Accuracy}} & \multirow{2}{*}{\textbf{Performance}} \\ \cline{2-7}
                                            & \textbf{INT}     & \textbf{FP}   & \textbf{INT}     & \textbf{FP}  & \textbf{INT}  & \textbf{FP}   &                                             &                      &                     &                                   &                                   \\ \hline
        \midrule
        \textbf{llm.npu}~\cite{xu2024fastondevicellminference} & INT4 & FP16/32      & /             & /     & INT8      & /          & \red{INT}                            &CPU \CPU{} $+$ NPU \NPU{}          &CPU \CPU{}                   & \red{Decrease (per dataset quantization)}  & \green{High}                        \\ \hline
        \textbf{Powerinfer2}~\cite{xue2024powerinfer} & /        & W4A16             & /             & /     & INT4      & W4A16      & \red{INT} \color{black}/ \green{FLOAT} &CPU \CPU{} $+$ NPU \NPU{}        &CPU \CPU{}                   & \red{Decrease (change model structure)}  & {\green{High}} / {\color{deepblue}Medium}                  \\ \hline
        \textbf{Qualcomm-AI}~\cite{Qualcomm-AI}       & INT4/8   & W4A16             & /             & FP16  & INT4/8    & /          & \red{INT}                            &NPU \NPU{}                         &NPU \NPU{}                   & \red{Decrease}                  & \green{High}                        \\ \hline
        \textbf{MLC}~\cite{mlc-llm}                   & /        & W4A16             & /             & W4A16 & /         & /          & /                                    &CPU \CPU{} $/$ GPU \GPU{}          &CPU \CPU{} $/$ GPU \GPU{}    & \green{Not affect}              & \red{Low}                                \\ \hline
        \textbf{Llama.cpp}~\cite{llamacpp}            & INT4/8   & W4A16             & /             & W4A16 & /         & /          & /                                    &CPU \CPU{} $/$ GPU \GPU{}          &CPU \CPU{} $/$ GPU \GPU{}    & \green{Not affect}              & \red{Low}                                 \\ \hline
        \textbf{Onnxruntime}~\cite{onnx}              & /        & FP16/32           & /             & /     & INT8/16   & /          & \red{INT}                            &CPU \CPU{} $/$ NPU \NPU{}          &CPU \CPU{} $/$ NPU \NPU{}    & \red{Decrease}                  & {\color{deepblue}Medium}                         \\ \hline
        \textbf{MNN-LLM}~\cite{mnn}                   & INT8     & W4A16             & /             & W4A16 & /         & /          & /                                    &CPU \CPU{} $/$ GPU \GPU{}          &CPU \CPU{} $/$ GPU \GPU{}    & \green{Not affect}              & {\color{deepblue}Medium}                                  \\ \hline
        \textbf{Ours}                                 & INT8     & W4A16             & INT8          & W4A16 & INT4/8    & W4A16      & \green{FLOAT}                        &GPU \GPU{} $+$ NPU \NPU{}          &GPU \GPU{} $+$ NPU \NPU{}    & \green{Not affect}              & \green{High}                    \\  
        \bottomrule
    \end{tabular}}
    \parbox[t]{\textwidth}{
        \vspace{5pt}
        \footnotesize If the framework supports multiple quantization methods, list only the common ones. "W4A16" indicates that weights are stored as INT4 and computations are performed in FP16.}
		\vspace{-5pt}
\end{table*}

\subsection{LLM Inference}
\label{sub:back:llm}

\CLSOSP{
Large Language Model (LLM) inference refers to the process of utilizing a pre-trained model to generate outputs based on user input.
It typically comprises two distinct phases: the prefill phase and the decoding phase.
In the prefill phase, the LLM processes the user’s input in a single batch to generate the first output token.
Given the potentially long input sequence, this phase relies on matrix multiplication operations, rendering it computationally intensive.
In contrast, decoding is a sequential and auto-regressive process, generating one token at a time.
It primarily involves matrix-vector multiplications, resulting in a memory-intensive workload.
}

% The token generated during the prefill phase serves as the starting point for the generation of the subsequent token. 
% This newly generated token then serves as input for the LLM, enabling the generation of the following tokens. 
% 

In contrast to cloud-side LLM inference (e.g. vLLMs~\cite{kwon2023efficient}, orca~\cite{yu2022orca}, etc.~\cite{serverlessllm, AlpaServe, MArk, MonoNN, DistServe, shubha2023adainf, Splitwise, Sarathi}), 
which prioritizes high throughput as well as meeting the response-time Service-Level Objectives (SLOs) of different inference workloads,
mobile-side LLM inference places a greater emphasis on minimizing end-to-end latency.
The latency can be further divided into two parts: TTFT (Time to First Token) and TPOT (Time per Output Token).
\CLSOSP{The former is primarily influenced by the speed of prefill phase processing, 
and the latter is associated with the token generation speed during the decoding phase.}
% The former, which denotes the latency until the generation of the first token, is primarily influenced by the speed of prefill phase processing. 
% The latter, which indicates the time required to produce each subsequent token, is associated with the token generation speed during the decoding phase.
% is intimately linked to the execution duration per iteration in decoding process.
% During the prefill phase, a batch of data needs to be processed. 
% On the other hand, thanks to the KV Cache(~\cite{CacheGen, InfiniGen, ainslie2023gqa}), the decoding phase processes only one token per time.
% Thus, TTFT is always constrained by computational bottlenecks, while TOPS is typically limited by bandwidth bottlenecks..

\begin{table}[b]
	\centering
    \setlength{\abovecaptionskip}{1pt}
    \setlength{\belowcaptionskip}{1pt}
    \vspace{-10pt}
	\caption{Specifications~\cite{nanoreview} of Mobile-side Heterogeneous SoC of mainstream vendors.}
	\label{tab:soc}
	\resizebox{\linewidth}{!}{
		\begin{tabular}{l|c|c|c|c|c|c}
			\toprule
			Vendor & SoC    & GPU & GPU FP16 & NPU & NPU INT8 & NPU FP16 \\ \hline
			\midrule
            Qualcomm & 8 Gen 3  & Adreno 750 & 2.8 TFlops  & Hexagon& 34 Tops & 17 TFlops \\ \hline
            MTK & K9300  & Mali-G720 & 4.0 TFlops & APU 790 & 48 Tops & 24 TFlops \\ \hline
            Apple & A18 & Bionic GPU & 1.8 TFlops & Neural Engine & 35 Tops & 17 TFlops \\ \hline
            Nvidia & Orin  & Ampere GPU & 10 TFlops & DLA & 87 Tops & None \\ \hline
            Tesla & FSD & FSD GPU & 0.6 TFlops & FSD D1 & 73 Tops & None \\  
			\bottomrule
		\end{tabular}
     }
     \caption*{\footnotesize NPU FP16: Since the vendors have not disclosed the computational power of the NPU for FP16, we roughly estimate it to be half of the INT8 computational power.}
\end{table}

\subsection{Mobile-side Heterogeneous SoC}
\label{sub:back:soc}

Considering the imperatives of personal privacy and security, there is a growing preference among 
individuals to deploy LLMs on local mobile devices instead of transmitting personal data to cloud services. 
Consequently, the mainstream vendors are actively enhancing their Edge-AI platform evolution, 
including mobile platforms such as Qualcomm's Snapdragon 8 Gen 3~\cite{Snapdragon8gen3}, Apple's A18~\cite{A18}, MediaTek's Dimensity 9300~\cite{A18}, Huawei's Kirin 9000~\cite{kirin9000}, etc.
Table~\ref{tab:soc} lists the parameter specifications of several mainstream mobile SoC platforms.
\CLSOSP{To meet the substantial computational demands of LLMs, mobile platforms are evolving towards heterogeneous SoC architectures.
In addition to traditional CPUs and GPUs, NPUs, which are dedicated processors optimized for ML workloads,
are increasingly playing a critical role due to their superior computational power compared to CPUs and GPUs.
Moreover, these heterogeneous processors often utilize a unified physical memory, distinguishing them from discrete heterogeneous systems.}
% which distinguishesthem from discrete heterogeneous systems
% To support the massive computational power required by LLMs, mobile platforms are evolving towards to the heterogeneous SoC. 
% In addition to the conventional CPUs and GPUs, NPUs are increasingly playing a critical role in these platforms.  Generally, these heterogeneous processing units can share a unified physical memory, which is significantly different from discrete heterogeneous systems.

% \FDH{It's worth noting that although vendors report high peak performance, the actual computational power 
% used by the main operators during LLM inference is significantly lower than the peak performance.} 
% On the Snapdragon 8 Gen 3 platform, we used QNN~\cite{QNN} to evaluate the actual computational power of the primary matrix multiplication in LLama3~\cite{dubey2024llama3}, as shown in Figure~\ref{fig:design-npu-perf}.

\subsection{Limitations of Mobile-side Inference Engines}
\label{sub:back:engine}

With the development of heterogeneous mobile SoCs,
numerous works have established frameworks for LLM inference to leverage the heterogeneous computational power at the edge,
such as llm.npu~\cite{xu2024fastondevicellminference}, PowerInfer2~\cite{xue2024powerinfer}, MNN-LLM~\cite{mnn}, MLC~\cite{mlc-llm}, etc~\cite{llamacpp, 234946, iyer2023automated,NCNN,TensorFlow-Lite,onnx}.
In Table~\ref{tab:comparison}, we provide a comprehensive comparison of existing approaches across multiple dimensions,
\FEHSOSP{including support for different backends, impact on model accuracy and end-to-end performance of LLM inference.}
% including support for different backends, impact on model accuracy, and the ability to utilize processors effectively at different stages of LLM inference.
Through this comparison, we identify two key aspects that existing works fail to address:
% \textemdash
% yet addressing these aspects holds the potential to significantly enhance the performance of on-device inference.

\myparagraph{Parallelizing computation between GPU and NPU.}
\CLSOSP{Although both GPUs and NPUs can accelerate the computation of LLM workloads and exhibit varying affinities for different operators,}
existing research primarily focuses on leveraging a single accelerator within mobile SoC.
For instance, MLC~\cite{mlc-llm} and MNN-LLM~\cite{mnn} exclusively utilize the GPU for computation, 
whereas PowerInfer~\cite{xue2024powerinfer} and llm.npu~\cite{xu2024fastondevicellminference} are limited to supporting the NPU backend. 
This challenge stems from the difficulty involved in synchronization and coordination among heterogeneous processors. 
Specifically, it is crucial to address how to efficiently synchronize data and effectively partition computational workloads across different backends.

% MLC~\cite{mlc-llm}, Llama.cpp~\cite{llamacpp}, and MNN-LLM~\cite{mnn} are all capable of leveraging GPU for computation.
% However, they lack effective mechanisms for coordinating and parallelizing workloads between the GPU and other processors.
% This challenge arises from the asynchronous nature of GPU execution, in contrast to the synchronous operations of CPU and NPU.
% Efficient parallelization thus requires precise coordination and data synchronization across heterogeneous processors.
% Moreover, since GPU kernels typically execute for only a few hundred microseconds,
% improper parallelization can lead to performance degradation rather than improvement.
% MLC~\cite{mlc-llm}, Llama.cpp~\cite{llamacpp}, and MNN-LLM~\cite{mnn} are all capable of leveraging GPU for computation.
% However, they lack effective mechanisms for coordinating and parallelizing workloads between the GPU and other processors.
% This challenge arises from the asynchronous nature of GPU execution, in contrast to the synchronous operations of CPU and NPU.
% Efficient parallelization thus requires precise coordination and data synchronization across heterogeneous processors.
% Moreover, since GPU kernels typically execute for only a few hundred microseconds,
% improper parallelization can lead to performance degradation rather than improvement.

\myparagraph{Fully utilizing NPU computational power without accuracy loss.}
\FEHSOSP{Being the most powerful computing unit on modern SoCs,
NPUs have been utilized by frameworks such as llm.npu~\cite{xu2024fastondevicellminference}, Qualcomm-AI~\cite{Qualcomm-AI} to accelerate the inference process.
However, these approaches primarily leverage the INT capabilities of NPUs for low-precision computations 
and do not fully exploit the available FLOAT capabilities. 
Utilizing INT4/8 for computations can lead to significant degradation in model accuracy (20\% reported by Qualcomm-AI), especially for small-sized edge models. 
Although llm.npu seeks to address this issue by identifying tensor outliers and employing mixed-precision techniques, 
this approach introduces sensitivity to the input dataset, 
resulting in inference accuracy that is highly dependent on the user's input. 
Previous work~\cite{xue2024powerinfer} also tries to utilize the FLOAT capabilities of NPUs but often yields inefficient performance. 
Due to the inherent hardware structure of NPUs, FLOAT performance can vary significantly based on tensor shape and ordering.}

%% file: charac.tex
\section{PERFORMANCE CHARACTERISTICS}
\label{sub:design:character}

% Contemporary mobile SoCs feature heterogeneous accelerators.
To effectively utilize the heterogeneous processors, 
we start by analyzing the performance of the GPU, NPU and memory system. 
These accelerators exhibit diverse performance characteristics stemming from their unique hardware architectures. 
In particular, NPUs display significant performance variability across different tensor types and operators.
Therefore, we conduct a comprehensive analysis of the architectural differences among these heterogeneous accelerators, 
and then summarize their performance characteristics.

\begin{figure}[htb]
    \setlength{\abovecaptionskip}{-3pt}
    \setlength{\belowcaptionskip}{-10pt}
    \includegraphics[width=\linewidth]{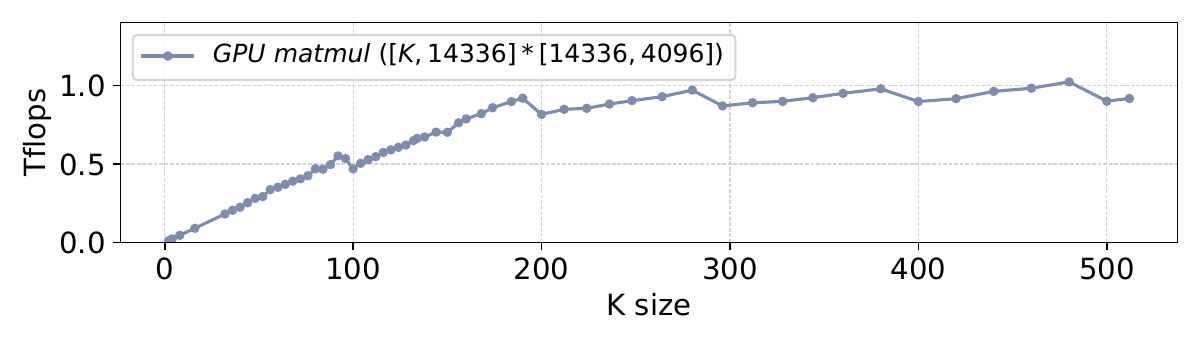}
    \caption{The GPU performance with varying tensor sizes.}
    \label{fig:design-gpu}
\end{figure}

\subsection{GPU Characteristics}
\label{subsub:design:GPU}

% Mobile GPUs share a similar computing architecture with discrete GPUs for the desktop, including SIMT instructions, on-chip shared memory and streaming multiprocessors or compute units (SM/CU). 
% The difference is that mobile GPUs employ a distinct memory hierarchy characterized by a unified memory address space (UMA) integrated into the system memory. 
% Operations like data transfers between CPU-side and GPU-side memory, which are necessary in discrete GPUs, become redundant in mobile GPUs. 
% However, traditional GPU frameworks such as OpenCL are not designed for these UMA-GPUs, 
% and adhere to the abstraction of a dedicated GPU memory for mobile GPUs.

\myparagraph{Characteristic GPU-1: Linear Performance.}
Figure~\ref{fig:design-gpu} illustrates the performance of mobile GPUs with varying tensor sizes. 
When the tensor size is small, GPU computation is memory-bound.
With the increase of tensor size, the total FLOPS increases linearly.
Once the size surpasses a certain threshold, GPU computation turns to be computation-bound,
and the total FLOPS stays stable.

\myparagraph{Characteristic GPU-2: High-cost Synchronization.}
There are two primary types of synchronization overheads associated with mobile GPUs. 
The first type arises from the data copy. 
Since existing GPU frameworks still maintain a separate memory space for mobile GPUs, 
developers must utilize APIs such as \texttt{clEnqueueWriteBuffer} to transfer data from the CPU-side buffer to GPU memory. 
\FEHSOSP{Secondly, the explicit synchronization command (\texttt{clFinish}) introduces a fixed latency of approximately 400 microseconds on our platform. 
This overhead results from blocking the execution of subsequent GPU kernels and synchronizing the states of the GPU driver. 
Upon completing the data transfer and executing the synchronization command, 
it ensures that all states are consistent between the CPU and GPU views, 
thereby enabling the consistent execution of subsequent CPU/GPU/NPU tasks.
}
% Unfortunately, this transfer process incurs a fixed latency, approximately 400 microseconds on our platform, regardless of data size. 
% The second type of overhead is related to kernel submission. 
% As GPUs adopt the asynchronous programming model, 
% subsequent kernels can be queued while the current one is executing, making the submission overhead negligible (about 10 to 20 microseconds). 
% However, after synchronization, the GPU queue becomes empty, 
% which causes an additional latency of 50 to 100 microseconds due to the overhead of kernel queueing and submission.

% There are two primary types of synchronization overheads associated with mobile GPUs. 
% The first type arises from the data copy. 
% Since existing GPU frameworks still maintain a separate memory space for mobile GPUs, 
% developers must utilize APIs such as \texttt{clEnqueueWriteBuffer} to transfer data from the CPU-side buffer to GPU memory. 
% Unfortunately, this transfer process incurs a fixed latency, approximately 400 microseconds on our platform, regardless of data size. 
% The second type of overhead is related to kernel submission. 
% As GPUs adopt the asynchronous programming model, 
% subsequent kernels can be queued while the current one is executing, making the submission overhead negligible (about 10 to 20 microseconds). 
% However, after synchronization, the GPU queue becomes empty, 
% which causes an additional latency of 50 to 100 microseconds due to the overhead of kernel queueing and submission.

\subsection{NPU Characteristics}
\label{subsub:design:NPU}

\begin{figure}[tp]
    \setlength{\abovecaptionskip}{8pt}
    \setlength{\belowcaptionskip}{-8pt}
    \includegraphics[width=\linewidth]{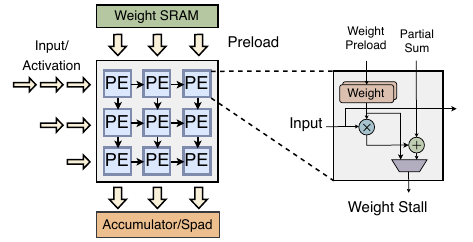}
    \caption{\textbf{The hardware design for NPU:} Systolic array with the weight stall computing paradigm.}
    \label{fig:design-npu-workflow}
\end{figure}

Although there are many different NPU implementations, 
matrix computation units (e.g., systolic arrays) serve as the most critical component inside NPUs. 
It leverages the data flow characteristics intrinsic to matrix computations, 
thereby minimizing the redundant load/store operations of model weights and activation. 
Figure~\ref{fig:design-npu-workflow} demonstrates a classical systolic array design.
In the computing flow of systolic array, weights are preloaded into each processing element (PE) prior to the computation. 
During the computation phase, a `weight stall' mode is employed, where weights remain stationary while inputs or activations are fed into the systolic array.
Finally, the computation results are output from the systolic array and are either stored in on-chip SRAM or directly forwarded to the subsequent systolic array unit.
Due to this NPU computation paradigm, NPUs exhibit three distinct computational characteristics: 
stage performance, order-sensitive performance and shape-sensitive performance.

\begin{figure}[htp]
    \setlength{\abovecaptionskip}{-3pt}
    \setlength{\belowcaptionskip}{-6pt}
    \includegraphics[width=\linewidth]{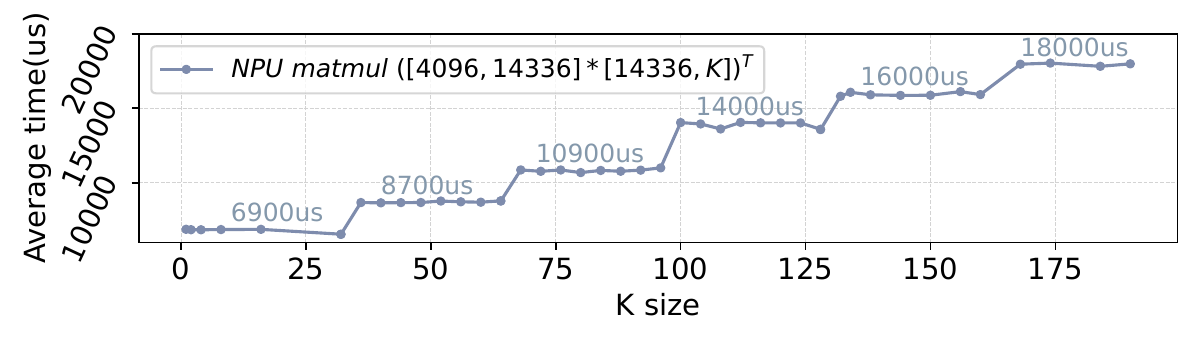}
    \caption{\textbf{The stage performance of NPUs.} The execution time of the Matmul operator across different tensor sizes.}
    \label{fig:design-npu-stage}
\end{figure}

\myparagraph{Characteristic NPU-1: Stage Performance. }
Due to the fixed size of the hardware systolic array within NPUs, 
the dimensions of the tensor used by the Matmul operator may not align with the size of the matrix computation unit,
which can lead to inefficient use of NPU computational resources.
As shown in Figure~\ref{fig:design-npu-stage}, 
this misalignment results in a phenomenon referred to as \emph{stage performance} across different tensor sizes.
\FEHSOSP{For instance, in the Snapdragon 8 Gen 3 SoC, NPU is equipped with multiple $32\times32$ systolic arrays.}
Thus, any computing tensor with dimensions smaller than $32$ will exhibit the same computational latency, 
leading to significant performance degradation for certain tensor shapes.
Furthermore, the compiler will partition tensors into tiles that align with the hardware size of the matrix computation unit. 
When tensor dimensions are not divisible by this size, 
the NPU compiler must introduce internal padding to align with the hardware requirement.
This alignment results in a stage performance effect during NPU calculations.

\begin{figure}[t]
    \setlength{\abovecaptionskip}{-1pt}
    \setlength{\belowcaptionskip}{-12pt}
    \includegraphics[width=\linewidth]{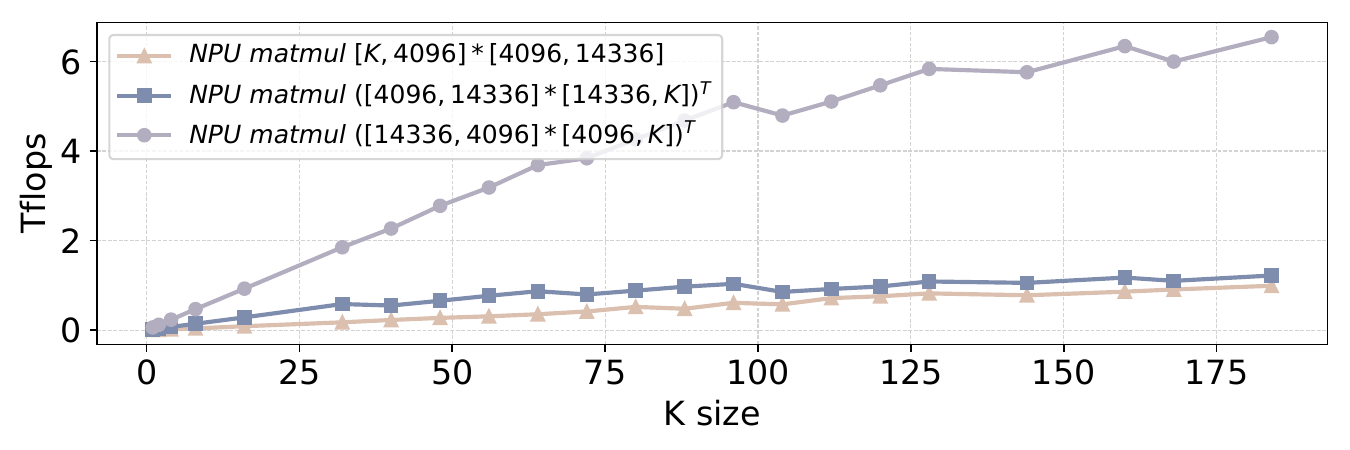}
    \caption{\textbf{The order-sensitive and shape-sensitive performance of NPUs.} The performance of the NPU is significantly influenced by the order and shape of the tensors.}
    \label{fig:design-npu-perf}
\end{figure}

\myparagraph{Characteristic NPU-2: Order-sensitive Performance.}
In addition to stage performance, NPUs also exhibit order-sensitive computation behavior. 
Consider two tensors with dimensions $[M, N]$ and $[N, K]$, where $M > N > K$. 
A conventional matrix multiplication (Matmul) operation requires $2 \times M \times N \times K$ operations.
If we reverse the order of these tensors, i.e., $[K, N] \times [N, M]$, the total number of computation operations remains unchanged. 
However, this can lead to significant performance degradation for the NPU, a phenomenon we refer to as \emph{order-sensitive performance}.
Figure~\ref{fig:design-npu-perf} presents a specific example where the matrix multiplication operation of $[14336, 4096] \times [4096, K]$ achieves 6$\times$ performance improvement compared to $[K, 4096] \times [4096, 14336]$.

The primary reason for order-sensitive performance is that NPU leverages the weight stall computing to minimize memory load/store overhead. 
In the ideal situation, the weight tensor always fits perfectly within the hardware matrix computation unit, eliminating the need for additional memory operations. 
However, when the weight tensor is significantly larger than the input tensor, it needs to load the weight tensor from the memory into the matrix computation unit more frequently, 
increasing memory overhead during the NPU execution.
As a result, although $[K, N] \times [N, M]$ and $[M, N] \times [N, K]$ involve the same number of computational operations, 
the larger size of $[N, M]$ results in inferior performance due to the extra memory operations involved.
In the worst-case scenario for NPU computation, if the weight matrix is infinite, 
the matrix computation unit cannot take advantage of the weight-stall computing paradigm,
and thus, the NPU performance may regresses to or even worse than the GPU performance. 

\myparagraph{Characteristic NPU-3: Shape-sensitive Performance.}
Similar to the order-sensitive performance, NPUs also exhibit \emph{shape-sensitive performance} characteristic. 
Even when the input tensor is larger than the weight tensor, the NPU's efficiency is influenced by the ratio between row and column sizes. 
More specifically, when the row size of the input tensor exceeds the column size, 
NPU demonstrates a better performance (compare the blue line with the purple line in Figure~\ref{fig:design-npu-perf}). 
Since the column size of the input tensor is shared with the weight tensor, a larger column size results in a larger weight tensor, 
which undermines the advantages of the weight-stall computation paradigm as we mentioned above.

% \begin{figure}[htp]
%     \setlength{\abovecaptionskip}{-1pt}
%     \setlength{\belowcaptionskip}{-15pt}
%     \includegraphics[width=\linewidth]{python_figs/memory_bandwidth.pdf}
%     \caption{\textbf{The total memory bandwidth with single and multiple processors.} 
%     We execute the decoding workloads across different backends and measure the available memory bandwidth in the entire SoC.}
%     \label{fig:design-mem-bandwidth}
% \end{figure}

\subsection{SoC Memory Bandwidth}
\label{subsub:design:memory}

\begin{figure}[t]
    \setlength{\abovecaptionskip}{-1pt}
    \setlength{\belowcaptionskip}{-10pt}
    \includegraphics[width=\linewidth]{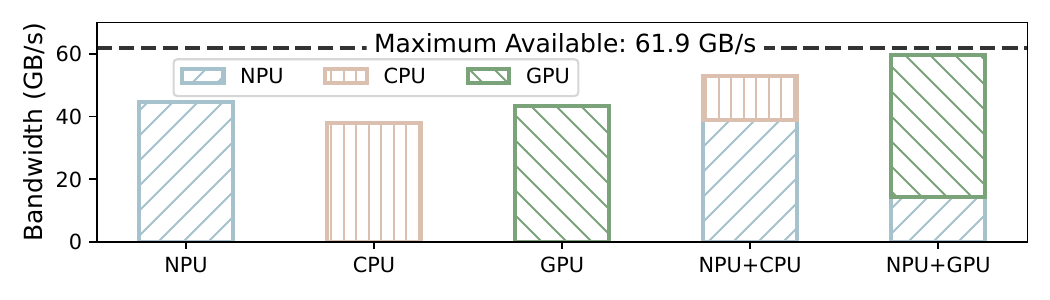}
    \caption{\textbf{The total memory bandwidth with single and multiple processors.} 
    We execute the decoding workloads across different backends and measure the available memory bandwidth in the entire SoC.}
    \label{fig:design-mem-bandwidth}
\end{figure}

\myparagraph{Characteristic Memory-1: Underutilized Memory Bandwidth with Single Processor.}
Although mobile SoCs adopt a unified memory address space across heterogeneous processors, 
our measurements indicate that no single processor can saturate the SoC's memory bandwidth during LLM decoding.
On the Snapdragon 8 Gen 3 platform, the theoretical peak memory bandwidth is 68 GB/s,
while the maximum achievable bandwidth is approximately 61.9 GB/s,
obtained only under continuous large-bulk memory operations
(e.g., multi-threaded memcpy or NEON-load intrinsics on the CPU, and vector or image-buffer loads on the GPU).
However, under decoding workloads, each individual processor (CPU, GPU, or NPU) achieves only 40–45 GB/s, as shown in Figure \ref{fig:design-mem-bandwidth}.
In contrast, concurrent GPU–NPU execution raises aggregate bandwidth utilization to around 60 GB/s,
with a 75/25 workload split that effectively overlaps GPU computation with NPU computation and synchronization overhead.
These results suggest that, as contemporary mobile SoCs are equipped with multiple memory channels,
co-execution of heterogeneous processors more fully utilizes the SoC's memory bandwidth,
creating a new opportunity to accelerate memory-bound LLM decoding.
% that GPU-NPU parallelism offers a new opportunity to improve LLM decoding performance,
% as the token generation rate is linearly correlated with the available memory bandwidth.

% \subsection(Cost Models for XPUs)
% To fully leverage the performance of the heterogeneous processors, 
% we first need to predict their performances with different tensor shapes.
% As tensor 

%% file: design.tex
\section{DESIGN}
\label{s:design}

\begin{figure*}[htb]
    \setlength{\abovecaptionskip}{5pt}
    \setlength{\belowcaptionskip}{-5pt}
    \includegraphics[width=\linewidth]{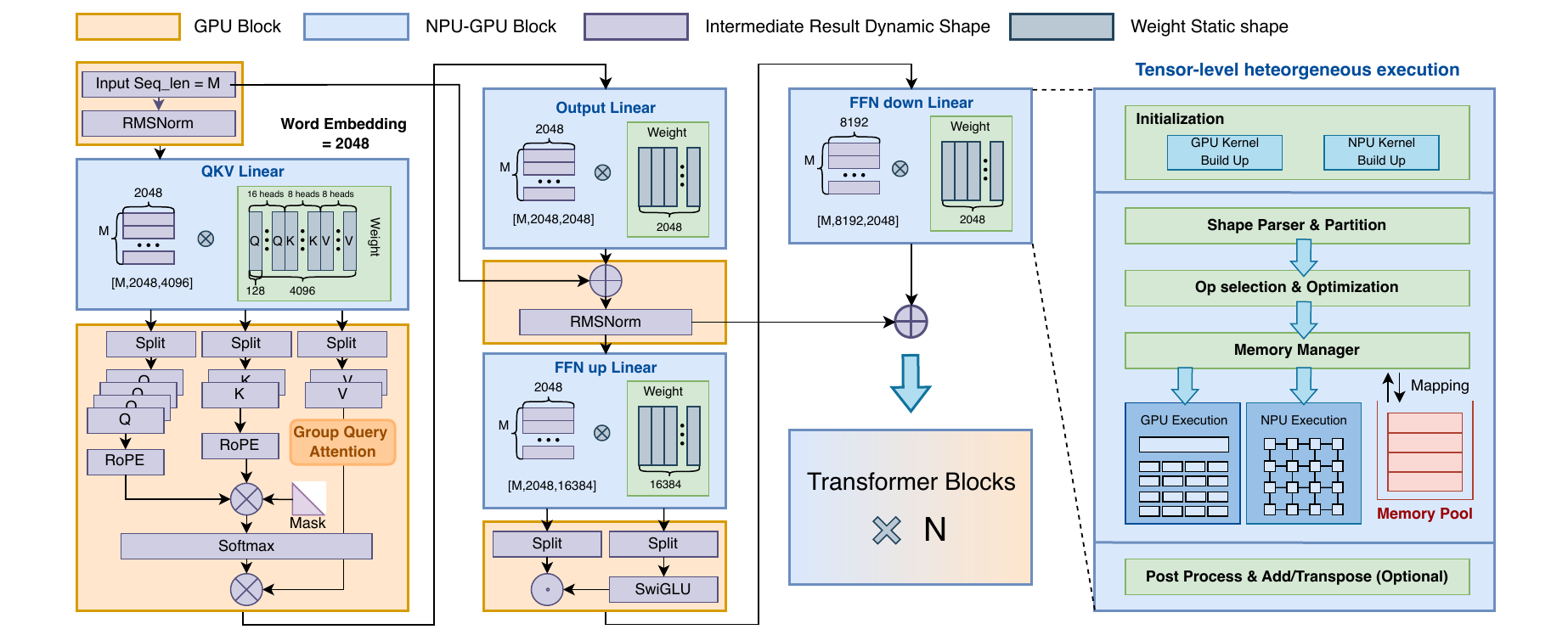}
    \caption{\textbf{The overall execution flow of a typical LLM with \sys:} Operators within the orange blocks are executed on the GPU backend, 
    whereas operators within the blue blocks can be offloaded to heterogeneous backends. 
    \sys additionally supports tensor-level heterogeneous execution through different tensor partitioning strategies.}
    \label{fig:design-llm_execution_flow}
\end{figure*}

% Given the power constraints and the presence of other applications in mobile systems, 
% we avoid exhausting all available power of heterogeneous processors for LLM tasks. 
% For instance, CPUs are ill-suited as dedicated backends for LLM tasks
% due to their low energy efficiency and engagement in general-purpose tasks.
% Consequently, \sys only utilizes the CPU as a control plane for synchronization, GPU kernel scheduling,
% and handling non-compute-intensive tasks such as dequantization.
% As for the others, the NPU outperforms the GPU in most cases but 
% can have significant performance degradation when doing certain calculations.
% Therefore, our system designates the NPU as the primary computing unit, 
% while leveraging the GPU to enhance the lower bound performance of the NPU in specific cases.

Given the distinct bottlenecks at different stages of LLM inference, our system adopts a stage-specific optimization strategy:
during the prefill stage, the objective is to maximize the computational throughput of the SoC,
whereas during the decoding stage, the focus shifts to maximize memory bandwidth utilization.
\CLSOSP{Figure~\ref{fig:design-llm_execution_flow} illustrates the overall execution flow of a typical LLM using \sys.}
Considering the power constraints of mobile systems and the need to accommodate other concurrently running applications,
we refrain from fully consuming all \FEHSOSP{available computing units} solely for LLM tasks.
% we refrain from fully consuming the all available power of heterogeneous processors solely for LLM tasks.
\FEHSOSP{For instance, CPUs, in particular, are not ideal as computational backends due to their relatively low energy efficiency and, 
more critically, their significant involvement in general-purpose processing.
Therefore, \sys leverages the CPU only as a control plane \CLSOSP{to handle tasks} such as synchronization and GPU kernel scheduling. }
% and lightweight operations such as dequantization.

\subsection{Layer-level GPU-NPU Execution}

We begin with a coarse-grained strategy that assigns computation tasks to the GPU and NPU based on \FEHSOSP{their computing affinities} at the layer level.
% Each operator is dispatched to the backend that is best suited for its execution.
For example, during the prefill phase, Matmul operators are allocated to the NPU,
which demonstrates superior performance in matrix multiplication computing operations.
In contrast, RMSNorm and SwiGLU operators are executed more efficiently on the GPU.
\FEHSOSP{Besides, given that LLM models typically have larger weight tensors compared to user input tensors, 
we adapt to the characteristics of \emph{NPU-2: order-sensitive performance} by exchanging the computation order of the input and the weight tensor, 
utilizing the following computational invariant:  $[M, N] \times [N, K] \rightarrow [[K, N] \times [N, M]]^T$.}
% the input and weight tensors are permuted from $[M, N] \times [N, K]$ to $[[K, N] \times [N, M]]^T$ prior to computation.
As for the decoding phase, due to \emph{NPU-1: stage performance}, the GPU becomes the primary computational unit,
as it offers better performance in matrix–vector operations.

% Figure~\ref{fig:design-llm_execution_flow} shows the overall execution flow of a typical LLM with \sys. 
% \sys employs two methods for GPU-NPU parallelism: \textbf{layer-level} heterogeneous execution and \textbf{tensor-level} heterogeneous execution. 
% The \textbf{layer-level} approach incorporates two key optimizations. 
% First, different operators are assigned to the most suitable backends:
% for instance, Matmul operators are directed to the NPU backend, 
% whereas RMSNorm/SwiGLU operators are more efficiently handled by the GPU backend. 
% Second, since typical LLM models feature larger size of weight tensors compared to user's input tensors,
% the input and weight tensors are permuted from $[M, N] \times [N, K] \rightarrow [[K, N] \times [N, M]]^T$,
% to meet \emph{NPU-\circled{2}: order-sensitive performance}.
% For the \textbf{tensor-level} approach, 
% \sys introduces various tensor partitioning strategies for different backends (\textsection\ref{sub:design:tensor-partition}), 
% and designs a solver to determine the optimal partition solution (\textsection\ref{sub:design:solver}).
% Both approaches adopt a novel synchronization technique to reduce synchronization overhead between the NPU and GPU (introduced in \textsection\ref{sub:design:fast-sync}).

\subsection{Tensor-level GPU-NPU Parallelism}
\label{sub:design:tensor-partition}

While the layer-level approach leverages both the GPU and the NPU to accelerate LLM inference,
it falls short of fully utilizing the capabilities of a heterogeneous SoC due to three key limitations:
\begin{myitemize}
    \item NPU performance degradation for specific tensor shapes.
    \item Underutilization of SoC memory bandwidth and computational power of heterogeneous processors.
    \item Static NPU graphs with high graph generation costs.
\end{myitemize}
To overcome these limitations, \sys introduces tensor-level parallel execution, enabled by three different partitioning strategies.

% \sys introduces the tensor-level heterogeneous execution with three distinct partitioning strategies:
% row-cutting, sequence-length cutting and hybrid-cutting. 
% These strategies address three key deficiencies associated with NPU-only execution: 
% (1) performance degradation for specific tensor shapes, 
% (2) static computation graphs with higher graph generation costs, 
% and (3) underutilization of the SoC memory bandwidth as well as computational power of heterogeneous processors.

\subsubsection{Weight-centric partition with static shape} 
\label{sub:design:tensor-partition-prefill}
\noindentparagraph{Problem.}
\FEHSOSP{When considering different tensor shapes, the NPU fails to outperform the GPU in all situations. 
This unexpected performance result arises from two factors: 
First, when the sequence length is short, the NPU cannot fully utilize all available computational resources due to \emph{NPU-1: stage performance}, 
resulting in performance that is similar to or even lower than the GPU's. 
Second, due to the dimensionality reduction matrix in the FFN-down layer, 
the column size of this matrix is larger than the row size (after transposition). 
This configuration is suboptimal for NPU execution, 
yielding only a 0.5$\times$ to 1.5$\times$ performance over the GPU, 
attributable to \emph{NPU-3: shape-sensitive performance}.
}

% Although the NPU can outperform the GPU by an order of magnitude in ideal scenarios, 
% its performance is significantly influenced by the shape of the weight and activation tensors.
% First, when the sequence length is short, NPU cannot exploit all available computational resources due to \emph{NPU-\circled{1}: stage performance}, 
% resulting in a similar or even lower performance compared to the GPU. 
% Second, due to the dimensionality reduction matrix inherent in the FFN-down, 
% the column size of this matrix is larger than the row size (after transposition). 
% This configuration is suboptimal for NPU execution even with large sequence lengths, owing to \emph{NPU-\circled{3}: shape-sensitive performance}. 
% In such scenarios, the NPU exhibits only 0.5$\times$ to 1.5$\times$ performance improvement over the GPU,
% as its computational efficiency on this tensor shape is extremely low.
% Notably, the MATMUL operation on this tensor alone accounts for nearly half of the total execution time during the prefill phase.

\begin{figure}[tp]
    \setlength{\abovecaptionskip}{8pt}
    \setlength{\belowcaptionskip}{-10pt}
    \includegraphics[width=\linewidth]{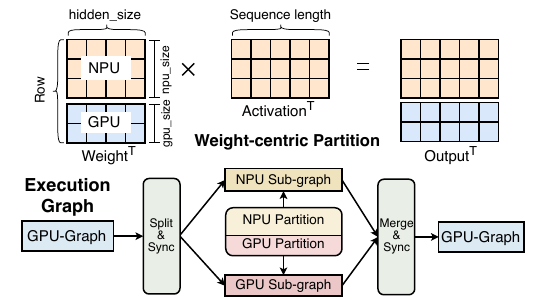}
    \caption{\sys partitions the weight tensor based on the row dimension, and dispatches computational loads to the different backends.}
    \label{fig:design-row-cutting}
\end{figure}

To address these performance limitations, we propose a \emph{weight-centric partitioning} strategy to enable parallel execution between the GPU and NPU.
\CLSOSP{To simplify the parallel strategy, we start by considering a fixed sequence length for user input,
enabling the NPU to construct the computation graph based on static tensor shapes.} 
As illustrated in Figure~\ref{fig:design-row-cutting},
this strategy partitions the weight tensor along its row dimension, assigning sub-tensors to the GPU and NPU for concurrent computation.
\CLSOSP{The partition ratio is statically determined by an offline solver (described in \textsection\ref{sub:design:solver}).}
% which allows the NPU computation graphs to be pre-generated with a fixed tensor shape.

\sys explicitly incorporates synchronization points into the computation graph and manages the splitting and merging of intermediate results across different backends. 
With an ideal partition ratio, the GPU and NPU complete their respective workloads simultaneously, thereby minimizing end-to-end inference latency.

Although the same partitioning strategy is applied, the rationale behind acceleration in the prefill and decoding phases differs.
During the prefill phase, the partitioning strategy leverages the GPU's computational resources to accelerate computation\textemdash
especially in scenarios where the NPU suffers from performance degradation due to unfavorable tensor shapes.
In contrast, during the decoding phase, the partitioning strategy is designed to address the underutilization of memory bandwidth when using a single computing unit (\emph{Memory-1}). 
This approach focuses on maximizing the memory bandwidth of the SoC while minimizing memory contention through GPU-NPU parallelism.

% To address these performance limitations, we propose a \emph{row-cutting} strategy for GPU and NPU parallelism.
% As illustrated in Figure~\ref{fig:design-row-cutting}, 
% row-cutting partitions the first tensor into sub-tensors based on the row dimension, 
% dispatching part of the computational workload from the NPU to the GPU. 
% To ensure that all activations (i.e., inputs) from the last layer are available before the next layer begins execution, 
% \sys explicitly incorporates synchronization points into the computation graph and manages the splitting and merging of intermediate results across different backends. 
% For an ideal partition, the GPU and NPU will complete their computations simultaneously, thereby reducing end-to-end latency.

\subsubsection{Activation-centric partition with dynamic shape}
\begin{figure}[tp]
    \setlength{\abovecaptionskip}{0pt}
    \setlength{\belowcaptionskip}{-15pt}
    \includegraphics[width=\linewidth]{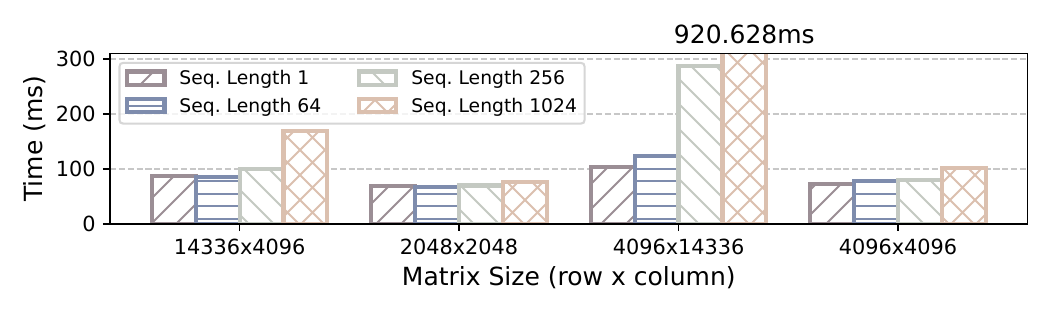}
    \caption{The NPU graph generation time for a single operator across various tensor shapes.}
    \label{fig:design-graph-generation}
\end{figure}

\noindentparagraph{Problem.}
\FEHSOSP{In real-world scenarios, user input typically has a dynamic sequence length, 
while current mobile-side NPUs only support static graph execution. 
This limitation arises from the dataflow graph compilation method~\cite{abadi2016tensorflow, team2016theano, tvm}, 
which is widely adopted by contemporary mobile NPUs~\cite{QNN, HIAI}. 
To address this issue, a straightforward solution is to generate the computation graph at runtime. 
However, this graph generation process incurs a non-trivial overhead that is proportional to the size of the tensors~\cite{ROLLER, Ansor, ragan2013halide}, as shown in Figure~\ref{fig:design-graph-generation}. 
In contrast, the GPU framework offers a set of kernel implementations, 
each of which can be adapted to accommodate a variety of tensor shapes. 
This capability facilitates dynamic-shape kernel execution at runtime.
}
% In addition to its fluctuating performance, 
% the mobile-side NPUs present another constraint: they only support static graph execution. 
% The shape and size of tensors at runtime need to be ascertained during the kernel initialization phase.
% This limitation stems from the dataflow graph compilation~\cite{abadi2016tensorflow, team2016theano, tvm}, a method widely adopted by current mobile NPUs~\cite{QNN, HIAI}. 
% Furthermore, as shown in Figure~\ref{fig:design-graph-generation}, the cost of graph generation for the NPU is highly dependent on tensor size, 
% as larger tensors expand the search space for optimization~\cite{ROLLER, Ansor, ragan2013halide}. 
% In contrast, the GPU framework provides a set of kernel implementations, each of which is adaptable to a variety of tensor shapes. 
% This facilitates the dynamic-shape kernel execution at runtime.

In order to support dynamic activation shapes on the NPU,
a common approach is to select a set of predefined tensor shapes\textemdash such as powers of two\textemdash and pad the activation tensors to match these shapes.
For example, if the input sequence length is 300,
the inference engine pads it to 512 to avoid the overhead of generating a new NPU graph for the unmatched tensor shape.
However, this padding introduces computational redundancy.

\begin{figure}[htp]
    \setlength{\abovecaptionskip}{2pt}
    \setlength{\belowcaptionskip}{-10pt}
    \includegraphics[width=\linewidth]{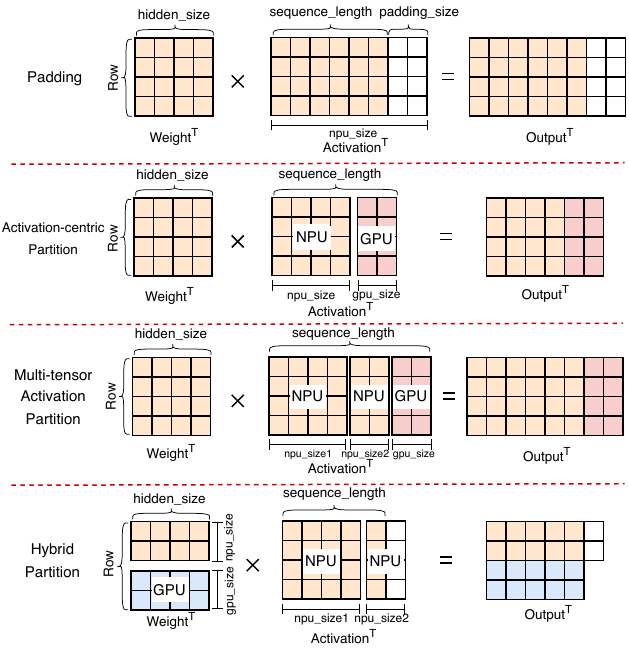}
    \caption{Due to the static computation graph of the NPU,
    \sys employs activation-centric partition to support the dynamic tensor shapes required by LLMs.}
    \label{fig:design-bs-cutting}
\end{figure}

To mitigate this issue, we propose an \emph{activation-centric partitioning} strategy that
enables support for dynamic tensor shapes while minimizing the overhead caused by excessive padding.
As shown in Figure~\ref{fig:design-bs-cutting}, 
\sys offloads computation tasks involving dynamic-shape tensors to the GPU (according to \emph{GPU-1: linear performance}), 
while retaining fixed-size tensor computations on the NPU. 
\FEHSOSP{For instance, consider an activation tensor with a sequence length of 300, 
which can be partitioned into two segments: 44 and 256. 
Since the segment of size 256 adheres to a standard shape, 
its computation graph is pre-generated and can be executed directly on the NPU. 
In contrast, the remaining segment of size 44 does not conform to a predefined shape for the NPU backend; 
therefore, it can be processed by the GPU backend in parallel with the NPU.}

% The remaining segment of size 44, which does not match any predefined shape, is processed by the GPU backend in parallel with the NPU.

% \myparagraph{Multi-tensor activation partition.}

\FEHSOSP{To further balance the computational load between the NPU and GPU, 
we adopt a \textbf{multi-tensor activation partitioning} strategy. 
Considering that the GPU generally exhibits lower performance than the NPU, 
we partition the activation tensor along the sequence length dimension into multiple sub-tensors, 
each conforming to a standard shape, along with one additional sub-tensor that has an arbitrary shape. 
All sub-tensors with standard shapes are executed sequentially on the NPU, 
while the sub-tensor with a dynamic shape is offloaded to the GPU. 
This approach allows us to minimize the computational load on the GPU and effectively balance the execution time between the GPU and NPU.
}

% Since the GPU generally exhibits lower performance compared to the NPU,
% its computations can become a bottleneck when the size of dynamic tensor surpasses a certain threshold. 
% To address this, we further partition the activation tensor along the sequence length dimension into multiple sub-tensors,
% each conforming to a predefined shape, along with one additional sub-tensor with an arbitrary shape. 
% All sub-tensors with predefined shapes are executed sequentially on the NPU,
% and the sub-tensor with a dynamic shape is offloaded to the GPU.

% For instance, if the input tensor's sequence length is 600, it can be divided into sub-tensors with sizes of 512, 32 and 56. 
% 512 and 32 are the pre-defined tensor shapes, which can be executed sequentially on the NPU, 
% while the sub-tensor with a dynamic size of 56 is offloaded to the GPU. 

\subsubsection{Hybrid partition}

% In addition to multi-sequence-length cutting, \sys can also employ a hybrid approach, which combines row-cutting and sequence-length cutting (\emph{hybrid-cutting}). 
% In this configuration, \sys continues to use padding for NPU computation while offloading a portion of the computational load to the GPU backend based on the row dimension.
% Through these elaborate tensor partition approaches, \sys can overlap the execution time between the GPU and NPU, 
% and further select the optimal partitioning strategies according to the different sequence lengths.
\noindentparagraph{Problem.}
\FEHSOSP{Although the activation-centric partitioning strategy facilitates support for dynamic tensor shapes, 
it results in suboptimal utilization of the computational resources of both the GPU and NPU due to \emph{NPU-3: shape-sensitive performance}. 
Specifically, the sub-tensor offloaded to the GPU may be either too large or too small to fully exploit its computational capabilities, 
or the tensor shape executed on the NPU may not be suitable for efficient NPU computation.}

\CLSOSP{To address this issue, we propose a \emph{hybrid partitioning} strategy that
combines activation-centric partition with pad\-ding to handle dynamic tensor shapes, 
while leveraging weight-centric partition to offload tensors to both GPU and NPU,
as illustrated in Figure~\ref{fig:design-bs-cutting}.}
\FEHSOSP{Given that the weight tensor is typically larger than the activation tensors, 
it allows for greater flexibility in partitioning the tensor into a more suitable shape for NPU computation. 
Consequently, the hybrid partitioning strategy not only maximizes the computational power of heterogeneous processors 
but also maintains system flexibility in accommodating dynamic user inputs.}

\subsection{Fast Synchronization}
\label{sub:design:fast-sync}
While GPU-NPU parallelism can reduce the execution time of certain operators, 
it may also introduce additional overhead due to synchronization between the GPU and NPU (\emph{GPU-2: high-cost synchronization}).
This issue arises because existing mechanisms, such as fences and \texttt{clFinish}, are not optimized for mobile SoCs.
The overhead becomes particularly pronounced during the decoding phase, 
where the execution time of the Matmul operator is reduced to only a few hundred microseconds. 
% \FEHSOSP{Specifically, each synchronization event incurs a latency of about 400 microseconds, 
% encompassing the blocking of subsequent GPU kernel launches and the synchronization of GPU driver states.}

% \CL{This overhead arises from two primary sources: first, the data copying between GPU memory and host memory (the current usage in OpenCL); 
% and secondly, the synchronization of GPU and NPU kernels that are originally intended for asynchronous execution.
% For instance, a naive approach might insert instructions like `clFinish' (for OpenCL) to ensure all GPU kernels are complete at synchronization points.}

To mitigate this overhead, we employ two strategies. 
First, mobile SoCs provide a unified address space, 
which allows mapping a memory buffer into both host and device address spaces, eliminating the need for additional data transfers. 
In the \sys runtime, a dedicated memory pool is reserved for allocating the input and output tensors of each operator. 
Since different layers in LLMs share the same decoder block, 
this memory pool requires only a few buffer slots, which can be reused across layers. 
Furthermore, these buffer slots will not be reclaimed by the \FEH{GPU / NPU} driver, 
ensuring that the mapping between CPU and \FEH{GPU / NPU} address spaces is maintained throughout model inference.

Second, we exploit the predictable waiting times for GPU kernels to facilitate fast synchronization. 
Given that LLMs execute identical operations across each layer, 
the waiting times for GPU kernels tend to be consistent and predictable across different layers.
\CL{We allow the synchronization thread to sleep for a predicted waiting time, followed by a polling mechanism to achieve precise synchronization.}
Since the minimum granularity of `usleep' in mobile SoCs is approximately 80 to 100 microseconds, 
it cannot serve as an accurate synchronization mechanism. 
Consequently, once the synchronization thread awakens, 
it utilizes a small/middle CPU core to continuously monitor the output tensor of the last layer. 
\CL{A flag bit is added alongside the output tensor and is updated once the output tensor is completely populated.}
The CPU core only needs to poll this flag bit for a few microseconds and can immediately notify the NPU for subsequent execution as soon as the GPU kernel completes.

\begin{figure}[tp]
    \setlength{\abovecaptionskip}{3pt}
    \setlength{\belowcaptionskip}{-10pt}
    \includegraphics[width=\linewidth]{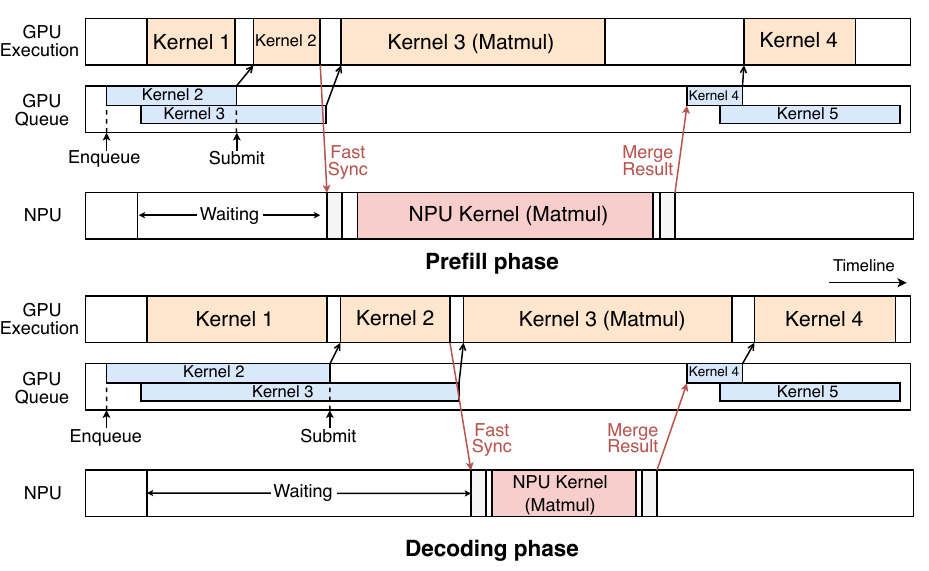}
    \caption{Fast synchronization during the prefill and decoding phases.}
    \label{fig:design-fasy-sync}
\end{figure}

While both the prefill and decoding phases leverage GPU and NPU parallelism with fast synchronization, 
several distinctions exist between these two phases, as shown in Figure~\ref{fig:design-fasy-sync}. 
In the prefill phase, the NPU exhibits superior computational capability, making it NPU-dominant. 
\sys effectively hides GPU execution time within NPU execution, but needs to delay the submission of the next GPU kernel until NPU execution is finished. 
Although this introduces a task submission overhead during GPU-NPU synchronization, 
this cost is approximately tens of microseconds and can thus be ignored in the prefill phase. 

Conversely, in the decoding phase, the GPU outperforms the NPU because GPU kernel implementations obtain more stable and higher memory bandwidth, making this phase GPU-dominant. 
Here, we overlap NPU execution with the GPU execution. 
\CLSOSP{Upon NPU task completion, the subsequent GPU kernel is promptly enqueued.
The GPU’s inherent queue ordering ensures correct synchronization for GPU kernels
without incurring additional submission overhead.
}

\subsection{Putting It All Together}
\label{sub:design:solver}

\begin{figure}[t]
    \setlength{\abovecaptionskip}{3pt}
    \setlength{\belowcaptionskip}{-13pt}
    \includegraphics[width=\linewidth]{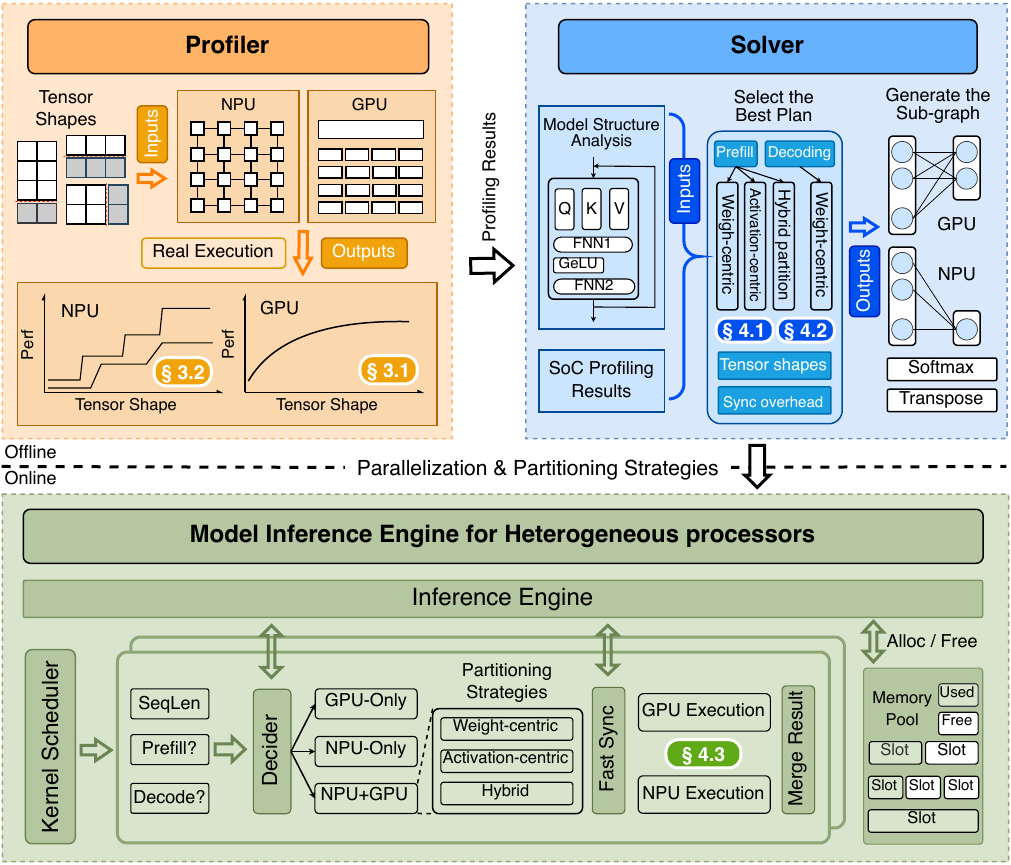}
    \caption{The overall architecture of \sys contains three components: Profiler, Solver and Inference Engine. }
    \label{fig:design-overall}
\end{figure}

Figure~\ref{fig:design-overall} presents the overall architecture of \sys.
Given a target SoC, the performance profiler first measures its performance matrices by executing operations on the actual NPU or GPU.
% The profiling results are then provided as input to the solver, along with the model architecture of a specified LLM.
% Based on these two inputs, the solver determines the optimal parallelization and partitioning strategies for running the given LLM on the specified SoC.
The profiling results are then fed to the solver, which determines how to schedule and partition the workload across processors for running a given LLM on the specified SoC.
The solver's output are subsequently used to generate the computation graph in advance. This entire process is performed offline.
During online inference, the inference engine dynamically selects an appropriate execution strategy based on the requested sequence length.

% Given a large language model, our tensor partition solver first identifies the various tensor shapes utilized by the model across several predefined sequence lengths. 
% It then cooperates with the performance profiler to determine the partitioning strategy of each tensor,
% and then partitions each tensor with the most suitable ratio. 
% Finally, it generates the computation graph for different backends, 
% conducting the execution of inference engine at runtime.

\myparagraph{Performance Profiler.}
\CLSOSP{The profiler executes the target operator with a variety of tensor shapes on both GPU and NPU to collect accurate performance metrics,
including execution time, memory bandwidth, and synchronization overhead.
The profiling space is constrained by three factors: (1) only weight tensor shapes from LLMs are considered;
(2) the NPU's stage performance characteristic impose minimum size requirements on sub-tensors;
and (3) activation tensors are restricted to a predefined set of standard sequence lengths.
Benefiting from this reduced set of tensor shapes, the profiling process can be completed efficiently, typically in less than 20 minutes.
}

% To determine the optimal partition solution, the solver works in conjunction with a performance profiler tailored for heterogeneous processors. 
% The profiler operates in two modes: real-execution and prediction. 
% In the real-execution mode, the profiler executes the target operator with various tensor shapes on actual hardware, 
% gathering precise performance metrics for both GPUs and NPUs. 
% Although this mode is time-consuming, it can be conducted offline. 
% \CL{Moreover, the NPU’s stage performance characteristic facilitates effective pruning of the tensor partition search space, 
% with constraints requiring row partitions to be aligned to 256 and sequence length partitions to 32, 
% thereby reducing the number of candidate partitions.}
% In addition to the real-execution mode, we provide a prediction mode. 
% Due to the inherent fluctuation in hardware performance, 
% \FDH{minor inaccuracies in performance results across different backends are tolerable for our solver.}
% Using traditional machine learning techniques, such as decision tree regression, we can predict NPU performance across different tensor shapes. 
% Conversely, given that GPU performance is more stable and less dependent on tensor shapes, 
% we easily estimate GPU execution time in compute-intensive scenarios using a fixed TFLOPS rate.

\myparagraph{Tensor Partitioning Solver.}
% \CLSOSP{The solver is responsible for determining the optimal partitioning strategy.
% It utilizes the profiling results and tensor shapes provided by the given model as input,
% and analyzes the execution time of different partitions along with the synchronization overhead and derive the ideal partitioning strategy and ratio.
% }
The solver determines the optimal tensor partitioning strategies for a given LLM based on profiling results.
It first analyzes the overall model structure to locate operators where workloads can be partitioned across heterogeneous processors,
such as those involved in attention projections and the FFN up/gate/down computations.
% which exhibit relatively higher computational demands.
For each operator, the solver enumerates all feasible parallelization strategies and
selects the one that best overlaps computation with synchronization to minimize total latency, as formulated in the following equation.
Since the prefill sequence length may vary arbitrarily while the profiler only provides measurements for standard sequence lengths,
the solver estimates operator latency for variable-length sequences by leveraging the performance characteristics of \emph{GPU-1: linear performance} and \emph{NPU-1: stage performance}.

\vspace{-10pt}
\[
\begin{aligned}
    T_{\text{total}} &= \min \Big( \max (T_{\text{GPU}}^{\text{partition1}}, T_{\text{NPU}}^{\text{partition2}}) + T_{\text{sync}} + T_{\text{copy}}, \\
              &\quad T_{\text{GPU}}^{\text{all}}, T_{\text{NPU}}^{\text{all}} + T_{\text{sync}} + T_{\text{copy}} \Big) \\
    &\quad \text{s.t.} \quad \text{Partition1} + \text{Partition2} = \text{All}.
\end{aligned}
\]
\vspace{-5pt}

Table~\ref{tab:solver} illustrates an example of the solver's inputs and outputs.
For tensor shapes in the decoding phase, a weight-centric partitioning strategy is adopted,
with the GPU performing the majority of computation.
This is because GPU usually outperforms NPU in the matrix-vector multiplication operation.
While for tensor shapes in the prefill phase, the optimal partitioning strategy is more variable and shape-dependent.
For instance, with a weight tensor of shape [4096, 4096], where a significant performance gap exists between GPU and NPU,
an activation-centric partition is applied when the input sequence length falls within the range of 257-272.
In contrast, for a weight tensor of shape [4096, 14336],
the computational capabilities of the NPU and GPU are relatively comparable (approximately 3:2),
primarily due to \emph{NPU-3: shape-sensitive performance}.
When the dynamic portion is relatively small\textemdash i.e., the prefill length only slightly exceeds a standard length\textemdash
activation-centric partition may result in under-utilization of GPU resources.
In such cases, a hybrid partitioning strategy is preferred.
% In contrast, for a weight tensor of shape [4096, 14336],
% activation-centric partitioning may lead to under-utilization of GPU resources when the dynamic portion is relatively small.
% In such cases, a hybrid partitioning strategy is preferred.

\begin{table}[t]
	\centering
    \setlength{\abovecaptionskip}{2pt}
    \setlength{\belowcaptionskip}{2pt}
	\caption{Input and output examples of the solver.}
	\label{tab:solver}
	\resizebox{\linewidth}{!}{
        \begin{tabular}{r|r|rr||c|c}
            \hline\hline
            \multirow{2}{*}{\textbf{\makecell{Weight\\Tensor Shape}}} & \multirow{2}{*}{\textbf{\makecell{Activation\\Tensor Shape}}} & \multicolumn{2}{c||}{\textbf{\makecell{Latency (us)}}} & \multirow{2}{*}{\textbf{\makecell{Partitioning Strategy}}} & \multirow{2}{*}{\textbf{\makecell{Partition Ratio\\(GPU : NPU)}}} \\ \cline{3-4}
            & & \multicolumn{1}{c|}{\textbf{GPU}} & \multicolumn{1}{c||}{\textbf{NPU}} & & \\ 
            \hline\hline
            4096, 4096    & 4096, 1           & \multicolumn{1}{r|}{511}   & \multicolumn{1}{r||}{693}   & Weight-centric     & 1 : 1                       \\ \hline
            28672, 4096   & 4096, 1           & \multicolumn{1}{r|}{1903}  & \multicolumn{1}{r||}{3886}  & Weight-centric     & 3 : 1                       \\ \hline
            4096, 14336   & 14336, 1          & \multicolumn{1}{r|}{1467}  & \multicolumn{1}{r||}{6506}  & No partition       & GPU-only                       \\ \hline
            4096, 4096    & 4096, 128         & \multicolumn{1}{r|}{7306} & \multicolumn{1}{r||}{912}  & No partition       & NPU-only                       \\ \hline
            4096, 4096    & 4096, [193-255]         & \multicolumn{1}{r|}{$9k\sim11k$}     & \multicolumn{1}{r||}{$1.2k\sim1.9k$}     & Padding & NPU-only                    \\ \hline
            4096, 4096    & 4096, 256         & \multicolumn{1}{r|}{10841} & \multicolumn{1}{r||}{1884}  & No partition       & NPU-only                       \\ \hline
            4096, 4096    & 4096, [257-272]         & \multicolumn{1}{r|}{$\sim11000$}     & \multicolumn{1}{r||}{$\sim1900$}     & Activation-centric & Dynamic : 256                    \\ \hline
            4096, 14336   & 14336, 256        & \multicolumn{1}{r|}{35231} & \multicolumn{1}{r||}{23445} & Weight-centric     & 2 : 3                       \\ \hline
            4096, 14336   & 14336, [257-384]        & \multicolumn{1}{r|}{$35k\sim50k$}     & \multicolumn{1}{r||}{$23k\sim32k$}     & Hybrid             & 2 : 3 (Weight)                       \\ \hline
        \end{tabular}
    }
    \caption*{\footnotesize Latency values for tensor shapes with dynamic ranges are estimated using the solver's latency function.}
    \vspace{-10pt}
\end{table}

% In the context of GPU-NPU parallelism, the solver traverses all possible partition solutions for the GPU and NPU, 
% and retrieves performance results from the profiler. 
% Commonly, the execution times for the GPU and NPU cannot be perfectly overlapped, 
% so the solver uses the maximum execution time of these two backends as the actual computation time. 
% Beyond computation time, the solver also accounts for synchronization overhead, 
% including kernel submission and activation transfer between GPU and NPU memory.
% The cost associated with kernel submission is further influenced by whether the execution is GPU-dominant or NPU-dominant.
% For certain tensor sizes where GPU-NPU parallelism does not yield any performance benefits, 
% the solver opts not to partition the tensor, instead selecting the optimal backend for execution.

\myparagraph{Inference Engine.}
During execution, a control plane decider determines whether a kernel is executed on the NPU backend, the GPU backend, 
or using GPU-NPU parallelism, based on \CLSOSP{the solver's output and current states}. 
When two adjacent kernels are allocated to different backends, the inference engine employs a fast synchronization mechanism to ensure data consistency. 
Upon completion of kernel execution on both backends, it merges intermediate results as needed. 
% In addition to scheduling and executing kernels on different backends, 
Besides, the inference engine manages a memory pool for host-device shared buffers, 
which are allocated or reclaimed as input and output tensors for each GPU/NPU kernel, 
bypassing the organization of the device driver.

% \begin{figure}[htp]
%     \setlength{\abovecaptionskip}{0pt}
%     \setlength{\belowcaptionskip}{0pt}
%     \includegraphics[width=\linewidth]{figs/GPU-NPU.pdf}
%     \caption{\textbf{Different computing patterns for GPU and NPU.}}
%     \label{fig:design-GPU-NPU}
% \end{figure}

%% file: eval.tex
\section{EVALUATION}

\begin{figure*}[htb]
    \setlength{\abovecaptionskip}{-3pt}
    \setlength{\belowcaptionskip}{-12pt}
    \includegraphics[width=\linewidth]{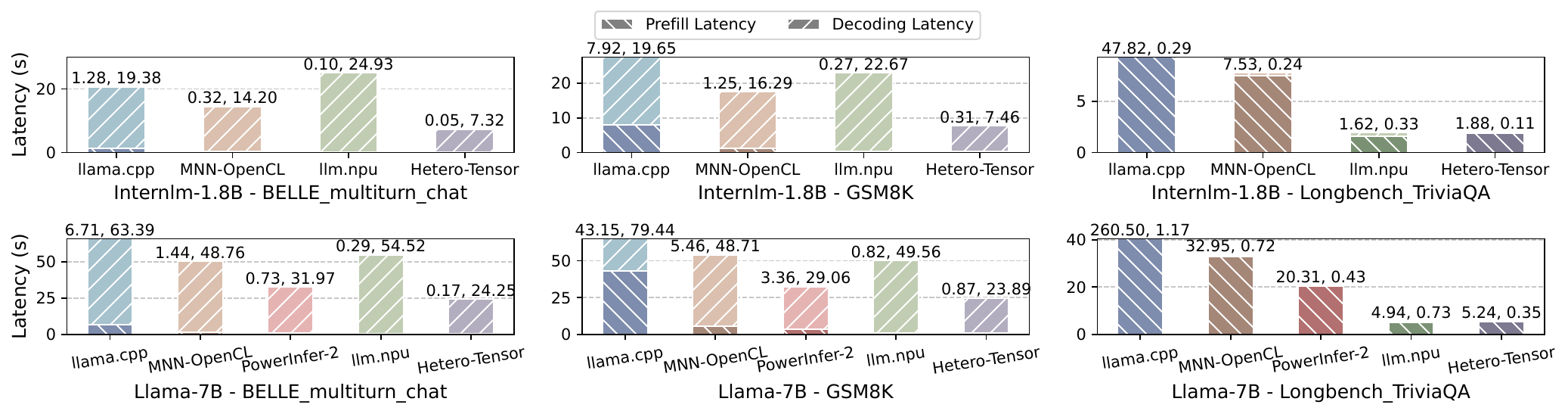}
    \caption{End-to-end latency comparison using real-world datasets on different models.
    The numbers labeled on each bar represent (prefill latency, decoding latency).}
    \label{fig:e2e}
\end{figure*}

\begin{figure*}[htb]
    \setlength{\abovecaptionskip}{0pt}
    \setlength{\belowcaptionskip}{-13pt}
    \includegraphics[width=\linewidth]{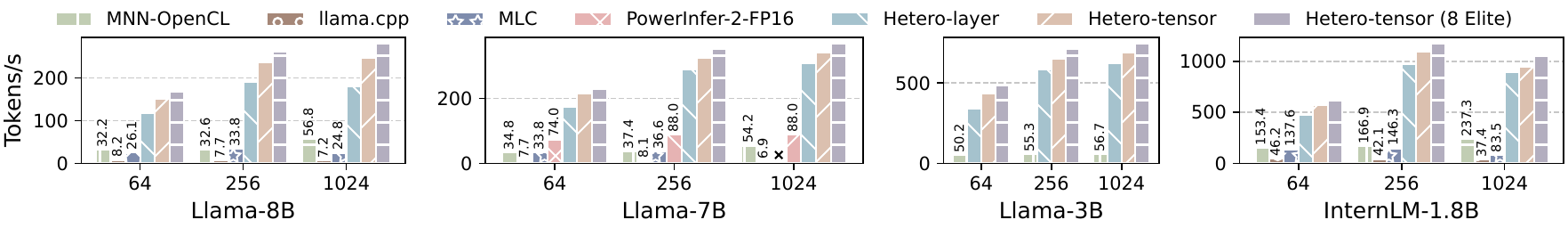}
    \caption{Prefill speed of MNN-OpenCL, llama.cpp, MLC, PowerInfer-2-FP16 and \sys
    across sequence lengths of 64, 256 and 1024 on multiple models.}
    \label{fig:prefill_perf}
\end{figure*}

\subsection{Experimental Setup}
% We have implemented a prototype of \sys based on PPL~\cite{PPL}, a SOTA mobile-side inference engine that supports both CPU and GPU. 
% \sys extends PPL by incorporating NPU support (using QNN-NPU library) and implements both layer-level and tensor-level heterogeneous execution across GPU and NPU.
We implemented an industrial-grade edge LLM engine: \sys, by developing optimized GPU kernels in OpenCL and incorporating NPU support via the QNN-NPU library.
The system supports both layer-level and tensor-level heterogeneous execution across the GPU and NPU.
As for the model quantization, 
\sys employs the W4A16 (weight-only) quantization\footnote{Only utilizes NPU's TOPS during the decoding phase, as NPU currently does not support W4A16 for decoding.}~\cite{frantar2022gptq, lin2024awq, li2024large, cheng2023teq, park2022lut} 
which balances the model accuracy (FLOAT computation) and storage overhead (INT4 for weight storage).
We evaluate the performance of \sys on both Snapdragon 8 Gen 3 and Snapdragon 8 Elite, two of the most powerful mobile SoCs.
For fair comparison with other works, all results in this paper are obtained on Snapdragon 8 Gen 3 unless otherwise noted.
Compared to previous approaches~\cite{xue2024powerinfer,xu2024fastondevicellminference,Qualcomm-AI} that leverage low-precision computation and exploit model sparsity,
\sys maintains the model accuracy and demonstrates even better performance through more efficient NPU utilization and GPU-NPU parallelism.

% \FEH{Even when compared to inference engines~\cite{xue2024powerinfer,xu2024fastondevicellminference,Qualcomm-AI} that exploit model/activation sparsity or rely on INT-only NPU calculations, 
% which can potentially compromise the model accuracy,
% \sys demonstrates comparable performance through more efficient NPU utilization and GPU-NPU parallelism, without sacrificing any accuracy.}
% Moreover, we believe that our techniques could also enhance performance for sparse inference, which remains an orthogonal aspect to our work.

In our evaluation, we compare \sys with representative mobile LLM inference engines that utilize GPU-only and NPU-only accelerators, 
including llama.cpp (CPU), MLC (GPU), MNN (GPU), llm.npu (NPU), and powerinfer-2 (NPU). 
We use batch size of 1 across all experiments, as concurrent requests are not yet common in current mobile-device deployment scenarios.
For end-to-end LLM workload tests conducted on the mobile platform, 
\sys demonstrates a performance improvement ranging from 1.34$\times$ to 6.02$\times$ across various workloads when compared to other SOTA solutions. 
More specifically, in the prefill stage, \sys achieves a performance enhancement of 3.29$\times$ to 24.9$\times$, 
whereas in the decoding stage, it attains a performance improvement of 1.50$\times$ to 2.53$\times$.
% \CL{\sys achieves significant performance improvements through layer-level heterogeneous execution (\htl) during the prefill phase, 
% with enhancements of 25.1$\times$, 7.27$\times$ and 3.18$\times$ compared to above inference engines respectively.} 
% Additionally, \sys gains a further 40\% improvement via tensor-level heterogeneous execution (\htt). 
% \FEH{During the decoding phase, \sys achieves a performance increase of up to 23.4\% by fully utilizing the SoC's memory bandwidth.}

\subsection{End-to-End Performance on Mobile Platform}

\begin{table}[b]
	\centering
    \setlength{\abovecaptionskip}{2pt}
    \setlength{\belowcaptionskip}{2pt}
    \vspace{-10pt}
	\caption{End-to-end benchmark configurations.}
	\label{tab:benchmark_config}
	\resizebox{\linewidth}{!}{
        \begin{tabular}{c|c|c|c}
        \toprule
        Task type & Dataset & Mean prefill tokens & Mean decoding tokens \\ \hline
        \midrule
        Multi-turn dialogue & BELLE multiturn chat & 54 & 374 \\ \hline
        Simple QA & GSM8K & 296 & 340 \\ \hline
        Long text processing & LongBench-TriviaQA & 1787 & 5 \\
        \bottomrule
        \end{tabular}
     }
    % \vspace{-10pt}
\end{table}

Figure~\ref{fig:e2e} compares the end-to-end latency of \htt with other edge LLM engines across three representative tasks:
multi-turn dialogue~\cite{BELLE}, simple QA~\cite{GSM8K}, and long-text processing~\cite{bai2024longbenchbilingualmultitaskbenchmark} (detailed configurations in Table~\ref{tab:benchmark_config}).
In the decoding-heavy multi-turn dialogue task with Llama-7B,
\htt achieves a 2.06$\times$ and 3.40$\times$ speedup over MNN and llm.npu, respectively,
and even outperforms PowerInfer-2 (with sparse model) by 1.34$\times$.
This is enabled by optimized GPU kernels and efficient GPU-NPU parallelism that fully utilize SoC memory bandwidth.
On GSM8K, where prefill and decoding are balanced, \htt yields a 2.62$\times$ average improvement.
In prefill-dominant workloads (Longbench), it delivers up to 6.02$\times$ speedup over MNN-OpenCL.
Notably, \htt still matches or surpasses llm.npu performance under such workloads,
despite the latter utilizes mixed-precision computation \CLSOSP{which requires per-dataset quantization and may cause accuracy loss.}

\FEHSOSP{To better analyze the performance of \sys, we conduct detailed tests of both prefill and decoding stages.}

\begin{figure*}[htb]
    \setlength{\abovecaptionskip}{-4pt}
    \setlength{\belowcaptionskip}{-10pt}
    \includegraphics[width=\linewidth]{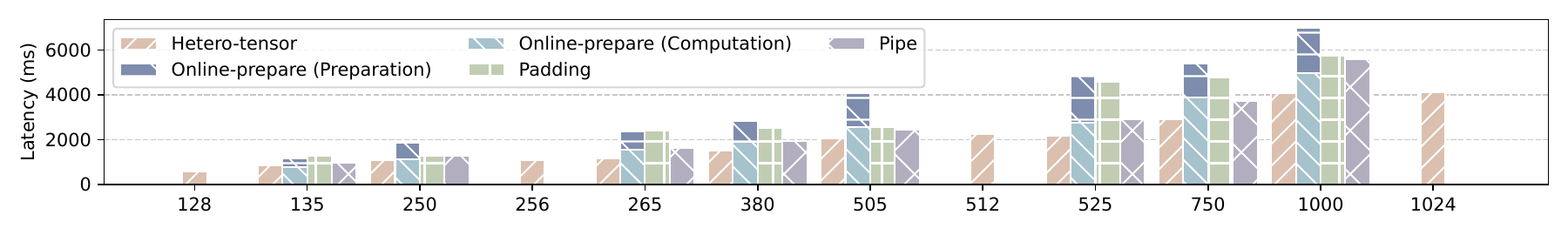}
    \caption{Prefill latency of Online-prepare, Padding, Pipe and \htt{} 
    on Llama-8B with dynamic sequence lengths.}
    \label{fig:Misaligned}
\end{figure*}

\subsection{Prefill Performance}

The evaluation of the prefill performance is conducted from two perspectives: 
fixed sequence lengths that fit in pre-defined graphs
and dynamic sequence lengths that do not align with the NPU's static graph. 

\subsubsection{\FEHSOSP{Fixed Sequence Length.}}

% The results show that \sys outperforms other frameworks under all models and prompt lengths.
% \htl can speed up 2.65$\times$ compared to ppl-OpenCL when running Llama-8B
% and \htt further improves the performance to 3.65$\times$.
Figure~\ref{fig:prefill_perf} compares the prefill performance across different frameworks with \FEHSOSP{several fixed sequence lengths}. 
For Llama-8B with sequence length of 256, 
\htl achieves 5.85$\times$, 5.64$\times$ and 24.9$\times$ speedup compared to MNN-OpenCL, MLC and llama.cpp.
% For sequence lengths 64 and 1024, \htl can speed up 3.67-14.36$\times$ and 3.17-25.12$\times$ respectively.
% These performance improvements result from utilizing the NPU as a substitute for the GPU in executing compute-intensive operators (e.g., Matmul), 
% given that a well-optimized NPU, with a more suitable tensor order, significantly outperforms the GPU in most scenarios.
Even compared to the NPU-only engine like PowerInfer-2-FP16, \htl achieves 3.29$\times$ speedup.
\FEHSOSP{These performance improvements arise from optimization mechanisms, 
including considerations of the affinity of NPU and GPU for different operators, 
as well as equivalent tensor order exchanges for Matmul computations.}
% thanks to a computation process meticulously optimized for the intrinsic characteristics of NPUs.

Based on this, \htt takes a step further and outperforms \htl by 30.2\% on average (up to 40.8\% when sequence length is 32).
% Under sequence length of 1024, \htt achieves 34.5$\times$, 9.99$\times$ and 4.36$\times$ performance improvement 
% over llama.cpp, MLC and MNN-OpenCL respectively.
The prefill speed of \htt reaches 247.9 tokens per second on Llama-8B
and is up to 1092 tokens per second on InternLM-1.8B. 
This is because \htt enables GPU-NPU parallelism by partitioning the weight and activation tensors (especially for FFN-down) into shapes that are suitable for NPU computation, 
while offloading the remaining part for GPU computation. 
This method further enhances the computational efficiency of both GPU and NPU.
On Snapdragon 8 Elite, \htt achieves about 10.5\% higher prefill performance than on Snapdragon 8 Gen 3,
benefiting from the enhanced capabilities of both the GPU and NPU.

Compared with other inference engines~\cite{xu2024fastondevicellminference,Qualcomm-AI} that only leverage NPU INT computation
and may sacrifice model accuracy (above 20\% accuracy degradation for Qualcomm-AI), 
\sys fully exploits the NPU's computational power (FLOAT).
Furthermore, with the efficient GPU-NPU collaboration, \sys achieves comparable or even better performance compared to these works.
For instance, \htt achieves 1092 tokens/s during the prefill phase with a sequence length of 256 \FDH{on InternLM-1.8B},
while llm.npu~\cite{xu2024fastondevicellminference} attains only 564 tokens/s under a similar model size.

\begin{figure}[b]
    \vspace{-10pt}
    \setlength{\abovecaptionskip}{0pt}
    \setlength{\belowcaptionskip}{0pt}
    \includegraphics[width=\linewidth]{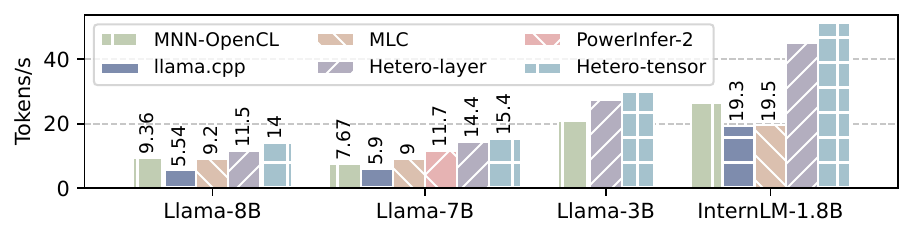}
    \caption{Decoding rate of MNN-OpenCL, llama.cpp, MLC, PowerInfer-2 and \htt on 
    different models. The prompt sequence length is 256.}
    \label{fig:decode_perf}
\end{figure}

\subsubsection{\textbf{Dynamic Sequence Length.}}

% \myparagraph{Baseline Selection.}
Since current mobile NPUs only support static computation graphs,
predefining a graph for every possible sequence length is infeasible.
A straightforward solution, termed \textit{Online-prepare}, generates a new NPU computation graph at runtime for each sequence length.
However, this approach is inefficient, as online graph preparation may cause significant performance degradation.

\FEHSOSP{Besides the straightforward solution, there are other alternative solutions such as \textit{Padding}, \textit{NPU-pipe} and \textit{Chunked-prefill}.
Padding mechanism preloads graphs of standard sizes (e.g., 128, 256, 512)
and pads the misaligned inputs to the nearest size.
NPU-pipe mechanism partitions a dynamic-size computation graph into multiple standard-size computation graphs (FFN layer) 
and executes these standard graphs sequentially on the NPU.
llm.npu adopts Chunked-prefill mechanism by splitting the input sequence into fixed-size chunks.
The chunk size must be chosen carefully to fully utilize the computational power of NPU and avoid unnecessary overhead.
For instance, Chunked-prefill can only achieve the maximum prefill performance until the sequence length is 1024,
and its performance is degraded to half when the sequence length is shortened to 256~\cite{xu2024fastondevicellminference}.}
% In contrast, Pipe (i.e., multi-sequence-length cutting without GPU support) chooses different standard sizes dynamically and has less overhead in graph loading.
% We adopt Pipe as the best solution for NPU-only prefill.}
% or partitioned using a multi-part segmentation strategy (see \textsection{\ref{sub:design:tensor-partition-prefill}}).
% The remaining segment, referred to as \textit{margin},
% is processed with a smaller graph and executed sequentially on the NPU.
% This method is referred to as \textit{Pipe}.
% MLLM-NPU splits sequence length into fixed size chunks (i.e., Chunked Prefill)
% to solve the problem of static computation graph on NPU.
% llm.npu addresses the challenge of static computation graphs on NPU 
% by splitting the input sequence into fixed-size chunks (referred to as "Chunked Prefill").
% The chunk size must be chosen carefully to fully utilize the computational power of NPU and avoid unnecessary overhead.
% For instance, Chunked Prefill can only achieve the maximum prefill performance until the sequence length is 1024,
% and its performance is degraded to half when the sequence length is shortened to 256~\cite{xu2024fastondevicellminference}.
% In contrast, Pipe (i.e., multi-sequence-length cutting without GPU support) chooses different standard sizes dynamically and has less overhead in graph loading.
% We adopt Pipe as the best solution for NPU-only prefill.}

% \myparagraph{Evaluation Results.}
Figure~\ref{fig:Misaligned} shows prefill latency of different methods under dynamic and misaligned sequence lengths.
Both graph preparation and computation times are reported for Online-prepare.
\htt consistently outperforms all baselines, achieving 2.24$\times$, 2.21$\times$, and 1.35$\times$ speedups
over Online-prepare, Padding, and NPU-pipe, respectively, at a sequence length of 525.
Online-prepare generally exhibits the highest latency due to graph preparation overhead,
which grows with sequence length and the number of NPU graphs (typically 4).
For instance, at length 135, preparation takes 408.4 ms (34.6\% of total latency), rising to 2050 ms at length 1000.
Padding causes stepwise increases in latency, leading to inefficiencies.
When sequence length slightly exceeds a standard size, 
the padding mechanism will introduce average 1.91$\times$ overhead compared with \htt.
NPU-pipe mitigates padding overhead by partitioning the whole computation graph into smaller subgraphs with standard sizes.
Compared to the NPU-pipe approach, \htt further reduces the prefill latency by 13.2\% to 30.1\%
by adaptively choosing the optimal partitioning strategy and enabling simultaneous execution of different computation subgraphs.
% \htt employs a more aggressive partitioning strategy that allows different computation subgraphs to be executed in parallel on both GPU and NPU. 
% Moreover, \htt{} can select different partitioning strategies for various sequence lengths based on the Tensor Partition Solver introduced in \textsection\ref{sub:design:solver}.

\subsection{Decoding Performance}

Figure~\ref{fig:decode_perf} presents the decoding speed of \htt and other inference engines.
\htt reaches up to 14.01 tokens/s on Llama-8B, 29.9 tokens/s on Llama-3B
and 51.12 token/s on InternLM-1.8B.
On Llama-8B, \htt achieves 1.50$\times$, 2.53$\times$ and 1.52$\times$ speedup 
compared to MNN-OpenCL, llama.cpp, MLC, respectively.
\CLSOSP{Even compared to PowerInfer-2 which utilizes the sparse model, \htt still achieves 1.32$\times$ speedup, 
because sparse computations resulting in numerous random and small memory accesses, compromising the overall memory bandwidth.} 
% On InternLM-1.8B, \htt achieves 1.94$\times$ speedup compared to MNN-OpenCL and 
% 2.62$\times$ speedup compared to MLC.
% \CL{As for \htl, since NPU computation is typically slower than GPU for small sequence length,
% it always chooses the GPU in decoding layers and performs similarly to PPL-OpenCL.}

\htt is the only framework that utilizes both GPU and NPU in the decoding phase, and maximizes the SoC memory bandwidth. 
When NPU and GPU are running concurrently, the memory bandwidth 
increases from 43.3 GB/s (only GPU) to 59.5 GB/s, achieving 96\% of the maximum available memory bandwidth.
\FEHSOSP{Given that the primary bottleneck during the decoding phase is memory bandwidth, 
we assert that \htt{} is approaching the theoretical maximum decoding performance.
Compared with \htl, \htt further leverages weight-centric partitioning strategy for the Matmul operator,
and achieves a 22.0\% speedup in decoding stage for Llama-8B model.}
We also evaluate \htt’s decoding performance on Snapdragon 8 Elite.
Since the memory bandwidth of our selected Snapdragon 8 Elite device is the same as that of Snapdragon 8 Gen 3 device,
the decoding performance remains similar.
% 8.52\% faster on Llama-3B and 13.38\% faster on InternLM-1.8B.

\begin{figure}[t]
    \setlength{\abovecaptionskip}{0pt}
    \setlength{\belowcaptionskip}{-15pt}
    \includegraphics[width=\linewidth]{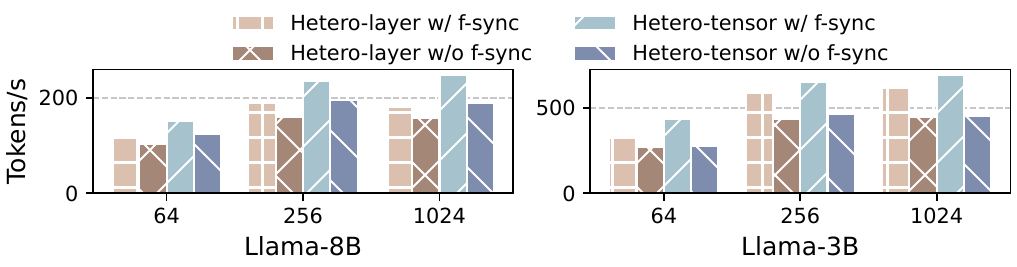}
    \caption{Prefill speed of \htl and \htt with and without fast synchronization 
    under sequence lengths of 64, 256 and 1024.}
    \label{fig:sync_prefill}
\end{figure}

\begin{figure}[t]
    \setlength{\abovecaptionskip}{-2pt}
    \setlength{\belowcaptionskip}{-10pt}
    \includegraphics[width=\linewidth]{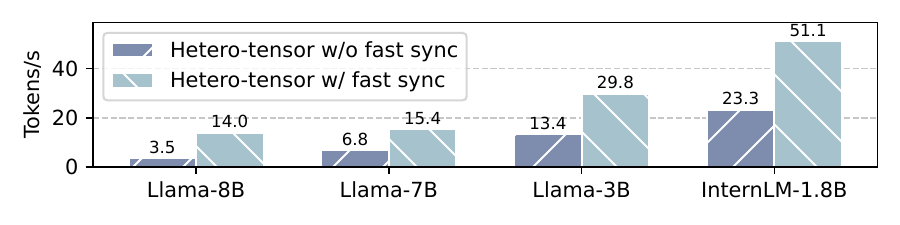}
    \caption{Decoding rate of \htt running Llama-8B, Llama-7B, Llama-3B and InternLM-1.8B 
    with and without fast synchronization. The prompt sequence length is 256.}
    \label{fig:sync_decode}
\end{figure}

\subsection{Effect of Fast Synchronization}
\label{sec:fast_sync}

\myparagraph{Prefill Performance.}
Figure~\ref{fig:sync_prefill} presents the prefill performance of \htl and \htt with and without fast synchronization.
On Llama-8B, fast synchronization yields average speedups of 15.8\% for \htl and 24.3\% for \htt.
Under a sequence length of 256, the prefill speed of \htt increases from 196.44 tokens/s to 236.92 tokens/s.
% For Llama-7B and InternLM-1.8B, 
% they achieve performance improvements of 49.0\% and 34.5\% with \htt and 31.7\% and 26.7\% with \htl, respectively.
% As \htt employs GPU-NPU parallelism within single-tensor computations (Matmul),
% it is more sensitive to the synchronization overhead, which can disrupt computational balance between the GPU and NPU.
Due to GPU-NPU parallelism, \htt is more sensitive to synchronization overhead,
which can disrupt computational balance between the GPU and NPU.

% On Llama-7B, the average improvements increases by 31.7\% and 49.0\%, similar to those of Llama-3B.
% On InternLM-1.8B, the average improvements are 26.7\% and 34.5\%.

% We can see that \htt gets more significant performance improvement than \htl 
% with the help of fast synchronization,
% because \htt carefully partitions the tensor between GPU and NPU to 
% make sure the time of GPU and NPU computation and  
% the time of GPU kernel queueing and submission can overlap with one another. 
% Without fast synchronization, unexpected synchronization overhead would break the balance.

\myparagraph{Decoding Performance.}
Figure~\ref{fig:sync_decode} presents the decoding performance of \htt
with and without fast synchronization.
% \htl is omitted because no synchronization is needed in its decoding phase.
On Llama-8B, the decoding rate of \htt speedups to 4.01$\times$ with fast synchronization.
On other models, we observe comparable improvements of 2.2$\times$ speedup.
The speedup in decoding phase is much higher than that in prefill phase,
because the execution time of each kernel in the decoding phase is much shorter.
Thus, the overhead of synchronization and GPU kernel submission is non-negligible.

\subsection{Ablation Study}

\begin{figure}[tb]
    \setlength{\abovecaptionskip}{0pt}
    \setlength{\belowcaptionskip}{-10pt}
    \includegraphics[width=\linewidth]{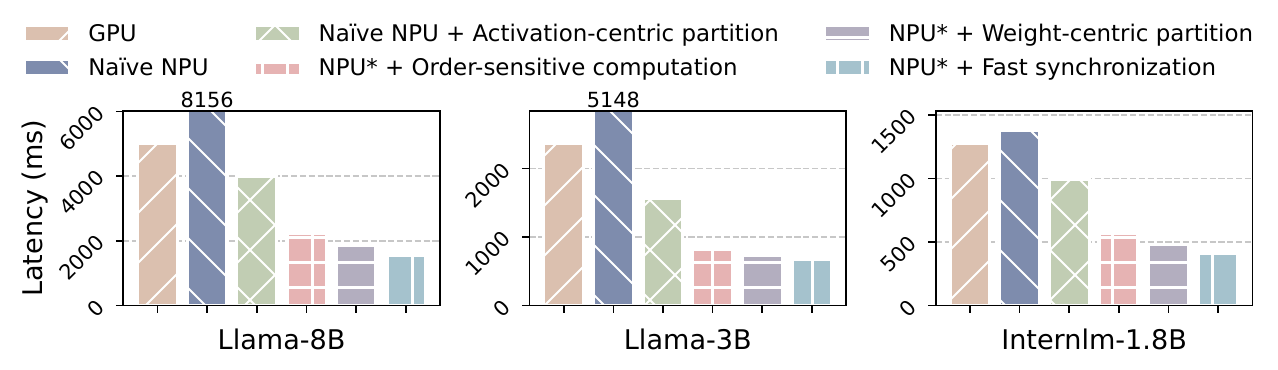}
    \caption{Ablation study of \sys when the prompt sequence length is 320.
    \emph{NPU*} stands for combination of all previous NPU techniques.}
    \label{fig:ablation}
\end{figure}

Since optimizations (NPU optimization, GPU-NPU parallelism, and fast synchronization) are only effective when applied jointly in the decoding stage,
our ablation analysis focuses on the prefill stage.
As illustrated in Figure~\ref{fig:ablation},
all of the techniques tailored to NPU and GPU characteristics effectively enhance prefill performance.
When running the Llama-8B model, naïve NPU implementation is even 62.5\% slower than GPU baseline due to the overhead of generating computation graphs on-the-fly.
Applying activation-centric partition eliminates this overhead, resulting in a 2.05$\times$ speedup.
\FEHSOSP{Building on this, \sys{} further achieves a 1.79$\times$ improvement by rearranging the tensor ordering during computation to address the order-sensitive nature of NPU. 
When considering the shape-sensitive performance characteristics of NPUs, 
enabling \CLSOSP{weight-centric partition} results in an additional 1.20$\times$ speedup. 
Finally, an efficient synchronization mechanism reduces GPU and NPU idle time during synchronization, 
leading to a further 1.19$\times$ speedup.}

% \textbf{Building on this, \sys further achieves a 1.79$\times$ improvement by exchanging the tensor ordering during computation to address the order-sensitive nature of NPUs.}
% % dynamically transposing matrices during computation to address the order-sensitive nature of NPUs.
% % Taking performance characteristics of both GPU and NPU into account, enabling GPU-NPU parallelism yields an additional 1.20$\times$ speedup.
% When taking the shape-sensitive performance characteristics of NPU into account, enabling GPU-NPU parallelism yields an additional 1.20$\times$ speedup.
% Finally, fast synchronization mechanism mitigates GPU and NPU idleness during synchronization, leading to a further 1.19$\times$ improvement.

\subsection{GPU Performance Interference}

To evaluate the impact of \sys on system-level image rendering as well as interference with other GPU applications, 
we conduct experiments by running GPU-only, \htl and \htt concurrently with a high-performance mobile game (League of Legends: Wild Rift),
all graphical settings in the game are kept at default values.

\begin{figure}[b]
    \vspace{-10pt}
    \setlength{\abovecaptionskip}{1pt}
    \setlength{\belowcaptionskip}{-2pt}
    \includegraphics[width=\linewidth]{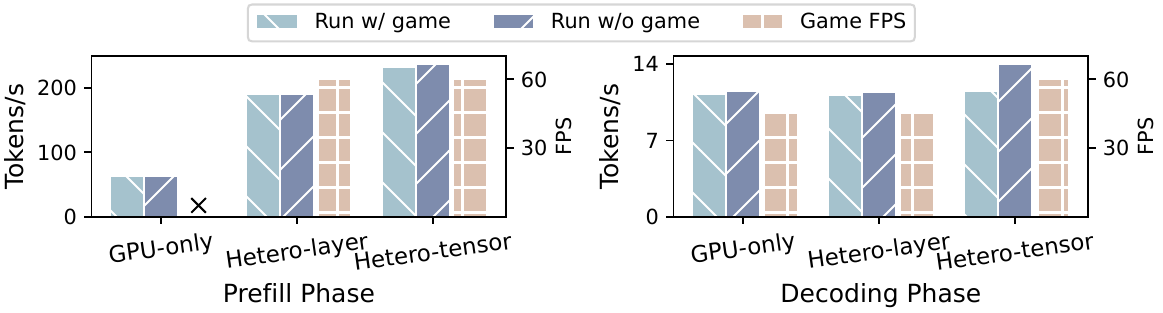}
    \caption{Prefill and decoding speed of GPU-only, \htl, and \htt with and without concurrent game execution, along with the game’s frame rate (FPS). Model: Llama-8B; prompt sequence length: 256.}
    \label{fig:game_test}
\end{figure}

As shown in Figure~\ref{fig:game_test}, when running concurrently with the game,
the prefill speeds of \htl and \htt decrease by only 0.5\% and 2.2\%, respectively,
while the game’s frame rate (FPS) remains unaffected.
In contrast, running concurrently with the GPU-only baseline results in a severe FPS drop to zero\textemdash
because the GPU kernels launched by the OpenCL runtime saturate the GPU submission queue,
preventing the game’s rendering tasks from completing within their required time budget.
In comparison, \htl and \htt offload only a small portion of computation to the GPU,
preserving sufficient GPU resources for timely game rendering.

In the decoding phase, the FPS drop of the GPU-only baseline is alleviated, with the game maintaining 46 FPS,
primarily because the GPU workload is lighter than in the prefill phase.
\htt continues to have no impact on FPS but suffers a 17.7\% reduction in decoding speed due to delayed GPU kernel execution,
which prevents full overlap between NPU and GPU execution.

\subsection{Energy Consumption}
% base_power = 0.295
% power = [4.53, 2.23, 2.87]
% # time of running 256 seqlen prefill
% elapsed_time = [4.024, 1.343, 1.080]
% energy = [18.22, 3.00, 3.10]

\begin{figure}[t]
    % \vspace{-10pt}
    \setlength{\abovecaptionskip}{0pt}
    \setlength{\belowcaptionskip}{-13pt}
    \includegraphics[width=\linewidth]{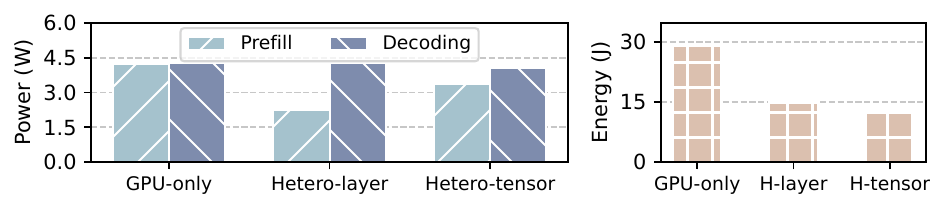}
    \caption{Power consumption during prefill and decoding, and end-to-end energy consumption for GPU-only, \htl, and \htt (prompt sequence length: 256; output length: 32).}
    \label{fig:energy}
\end{figure}

Figure~\ref{fig:energy} presents the power and energy consumption of GPU-only, \htl, and \htt.
During the prefill phase, \htl exhibits the lowest power consumption of 2.23$\mathrm{W}$,
primarily due to its reliance on the NPU for most computations.
In the decoding phase, \htt achieves the lowest power consumption by offloading computation to both the NPU and GPU.
In terms of end-to-end energy consumption, \htt is the most efficient,
consuming 55\% less energy than GPU-only and 12.8\% less than \htl, owing to its faster execution.

%% file: discuss.tex
\section{DISCUSSION}
\label{s:discussion}

\myparagraph{Platform and Model Generality.}
% \sys is broadly generalizable to both other mobile platforms and diverse model architectures.
Modern mobile SoCs commonly adopt similar hardware characteristics, such as unified memory architecture,
an asynchronous GPU execution model, and a systolic-array-based NPU.
Since the core observations and techniques of \sys are designed around these widely adopted features,
the system can be ported to other mobile SoCs with minimal retuning.
In terms of different models, even when the model architecture changes (e.g., MoE),
the underlying operator types invoked during computation remain largely consistent.
Our design focuses on operator-level optimization and enables GPU-NPU parallelism at the kernel granularity.
% The profiler and solver are designed to be adaptable to different matrix sizes.
As a result, the system is also applicable to models with diverse architectures and parameter configurations.

\myparagraph{Model Quantization.}
\sys employs W4A16 (weight-only) quantization, striking a balance between memory footprint and computational accuracy. 
W4A16 quantization is the most widely used approach in real-world deployments. 
W4A16 quantizes model weights during storage and dequantizes them to FLOAT for computation. 
In contrast, other approaches~\cite{xu2024fastondevicellminference,xue2024powerinfer,Qualcomm-AI} require quantization of both activations and weights 
to INT, which impairs inference accuracy.

\myparagraph{Parallelism among Accelerators.}
While the parallelization of ML workloads is a well-established field,
existing research\cite{serverlessllm, yu2022orca} has predominantly focused on distributing computations across homogeneous, discrete accelerators like NVIDIA GPUs and Google TPUs.
Some prior approaches\cite{song2020accpar,jia2022whale}, such as AccPar, have addressed heterogeneous systems.
However, their scope is often confined to cloud-based scenarios,
and the accelerators considered are typically from the same vendor lineage (e.g., TPUv2 and TPUv3),
which feature similar hardware architectures and a unified software stack.
In contrast, our work addresses the challenge of parallelization across fundamentally distinct accelerators integrated within a single mobile SoC.

\myparagraph{Design of Future Edge AI Accelerators and Systems.}
Based on our experience in optimizing LLMs on commercial SoCs,
we suggest that current edge AI SoCs can be further improved along following dimensions.
First, unified GPU–NPU scheduling is needed. As future edge scenarios will involve increasingly complex multi-task workloads,
including GPU-only, NPU-only, and GPU–NPU parallel tasks\cite{shen2025xsched},
the absence of a unified scheduling mechanism may lead to situations where tasks running on one processor simultaneously block the execution of tasks on both processors.
Second, although current mobile SoCs adopt UMA, they often lack coherent memory management across different processing units.
In our implementation, we establish shared memory between the CPU and GPU using OpenCL,
between the CPU and NPU via QNN APIs, and further enable GPU-NPU shared memory.
A unified API and memory management layer across CPU, GPU, and NPU would greatly simplify development and reduce unnecessary synchronization overhead.
Third, a fast and lightweight synchronization library for heterogeneous processors would facilitate efficient overlap of computation and communication.

% \myparagraph{Shared memory between GPUs and NPUs.}
% Current mobile SoCs (e.g., Apple M/A series, Qualcomm Snapdragon series) support a unified address space for CPU, GPU and NPU. 
% \FEH{In our implementation, we have successfully established shared memory between the CPU and GPU using OpenCL, 
% and between the CPU and NPU using QNN APIs. 
% Additionally, by employing the \texttt{CL\_MEM\_USE\_HOST\_PTR} flag, 
% we can also map the NPU's shared memory to the GPU, thereby establishing shared memory between the GPU and NPU.
% We recommend that mobile SoC vendors provide native APIs to allocate unified memory for all heterogeneous processors.}

% and shared memory between the CPU and NPU.
% Furthermore, we can also map the }
% However, since the Qualcomm's NPU driver is close-sourced, 
% we are unable to modify it to support shared memory with the GPU. 
% Consequently, we must store portions of the model weights in both NPU and GPU memory, 
% as tensor partitioning may vary between the prefill and decoding phases. 
% This is not a limitation of \sys itself but rather a temporary restriction due to the current Qualcomm NPU driver implementation.

%% file: concl.tex
\section{CONCLUSION}
\label{s:concl}
This paper introduces \sys, the fastest LLM inference engine optimized for heterogeneous processors in modern mobile SoCs. 
We conduct an in-depth analysis of the GPU and NPU hardware architectures, highlighting the NPU's tensor shape-sensitive performance characteristics. 
\sys leverages the GPU to improve the lower bound of NPU performance and enhance computational flexibility.
It introduces several key techniques, including tensor partition, fast synchronization, and profiler-solver collaboration,
to address the challenges of underutilized memory bandwidth and computing resources.
\sys demonstrates a 1.34$\times$ to 6.02$\times$ speedup in real-world benchmarks,
while maintaining negligible interference with other applications.
Finally, this work offers new insights into the design of more efficient edge AI accelerators and heterogeneous SoCs.